\documentclass[12pt,fleqn]{tsp}
\usepackage{url}
\usepackage{tikz}
\usepackage{calc}
\usepackage{amsthm}
\usepackage{enumitem}
\usepackage{setspace}
\usepackage{multirow}
\usepackage{mathptmx}
\usepackage{mdframed}
\usepackage{fancyhdr}
\usepackage{tcolorbox}
\usepackage{tcolorbox}
\usepackage{tabularx,colortbl}
\usepackage[hang,flushmargin]{footmisc}
\techscience{vol}{no}{2025}{0}
\license{\includegraphics[width=16cm,height=1.5cm]{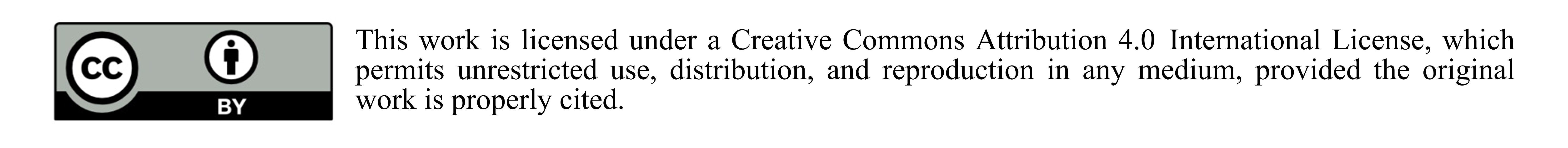}}
\thehead{Computer Modeling in Engineering {\&} Sciences}%
\newtheoremstyle{note}
{3pt}
{3pt}
{\normalfont}
{}
{\bfseries}
{.}
{.5em}
{}
\theoremstyle{note}

\graphicspath{{./image/}}
\runningtitle{Shock-capturing particle hydrodynamics with reproducing kernels} 
\title{Shock-capturing particle hydrodynamics with reproducing kernels}
\author{
  Stephan Rosswog{\thanks{University of Hamburg, Hamburg Observatory, Gojenbergsweg 112, 21029 Hamburg, Germany.}$^{\bm ,}$\thanks{The Oskar Klein Centre, Department of Astronomy, AlbaNova, Stockholm University, SE-106 91 Stockholm, Sweden.}$^{\bm ,}$$^\ast$
  \vskip 0.15cm
  \centerline{$^\ast$Email: stephan.rosswog@uni-hamburg.de}
  \vskip 0.15cm
  \centerline{Received: Month Day, Year; Accepted: Month Day, Year.}}
}

\newcommand{\Ma}{\texttt{MAGMA2}\;}

\def\p{\partial}

\def\be{\begin{equation}}
\def\ee{\end{equation}}
\def\bi{\begin{itemize}}

\def\ei{\end{itemize}}
\def\ben{\begin{enumerate}}
\def\een{\end{enumerate}}
\def\bea{\begin{eqnarray}}
\def\eea{\end{eqnarray}}
\def\ra{\vec{r}_a}
\def\rb{\vec{r}_b}

\begin{document}
\maketitle

\begin{abstract}
\addbottompattern
\begin{mdframed}[backgroundcolor=gray!8,linewidth=0pt,nobreak=true,innerleftmargin=2.5cm,innerrightmargin=2.5cm,leftmargin=-2cm,rightmargin=-2cm,]

We present and explore a new shock-capturing particle hydrodynamics approach. 
Our starting point is a commonly used discretization of smoothed particle hydrodynamics. We enhance this discretization with Roe's approximate 
Riemann solver, we identify its dissipative terms, and in these terms, we use slope-limited linear reconstruction. All gradients needed for our method 
are calculated with linearly reproducing kernels that are constructed to enforce the two lowest-order consistency relations. We scrutinize our reproducing 
kernel implementation carefully on a "glass-like" particle distribution, and we find that constant and linear functions are recovered to machine precision. 
We probe our method in a series of challenging 3D benchmark problems ranging from shocks over instabilities to Schulz-Rinne-type vorticity-creating shocks. 
All of our simulations show excellent agreement with analytic/reference solutions.

\end{mdframed}
\end{abstract}

\section{Introduction}
\label{sec:intro}
\begin{figure} 
   \centering
   \centerline{
   \includegraphics[width=0.5\columnwidth]{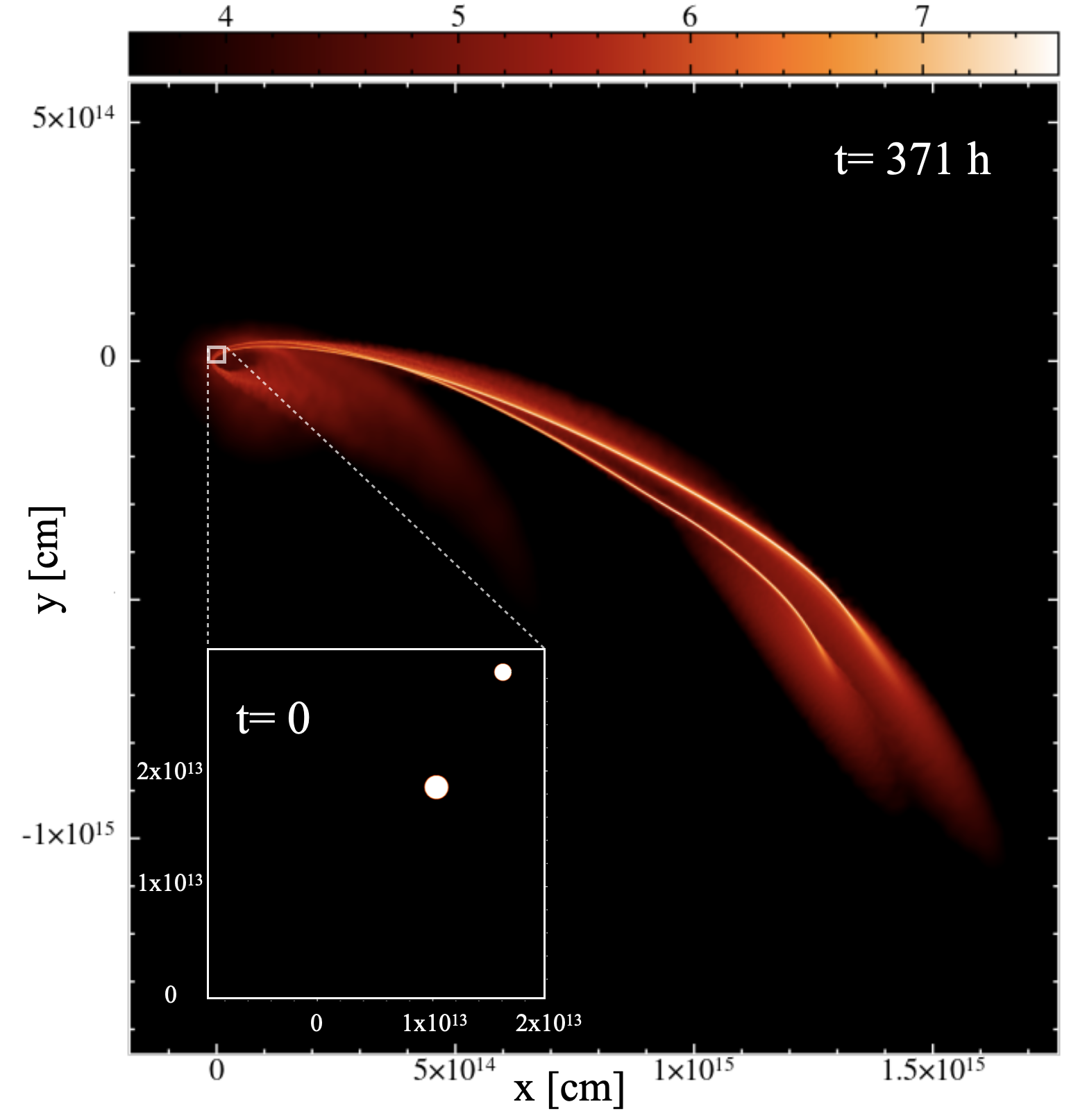}}
    \caption{Illustration of one of SPH's most salient features: the natural treatment of vacuum. The plot shows how two stars have been ripped apart by a 
    black hole (lurking at the coordinate origin) into spaghetti-like, thin gas streams 
    that are held together by the gas' self-gravity. The inset shows the initial conditions of the initially spherical stars. The original
    stars in the inset only cover a fraction of $10^{-9}$ of the volume that is shown at late times (t= 371 h). Simulation performed with the \Ma code, see Rosswog (2020).}
   \label{fig:motivation}
\end{figure}

The Smoothed Particle Hydrodynamics (SPH) method was originally developed to solve astrophysical gas dynamics problems \cite{lucy77,gingold77}. 
In an SPH simulation, the particle distribution automatically adapts to the dynamics of the gas flow, and this has major advantages when, for example,  
modeling the gravitational collapse of gas clouds that condense into denser filaments and finally form stars. The natural geometric adaptivity of SPH 
also has advantages in many other contexts where challenging geometries are involved, e.g, in simulating dam breaks, e.g. \cite{xu21}, or fracture processes,
e.g. \cite{rahimi22}, see \cite{monaghan12a} for a broad range of SPH applications.  Since the particles 
automatically follow the gas flow, this also implies that vacuum is simply modeled by the absence of computational particles. Eulerian methods, 
in contrast, need to model vacuum as a low-density background gas, and this can cost substantial
computational resources even though one is not interested (and nothing is physically going on) in  empty space. As an admittedly extreme, but 
astrophysically relevant example, we show in Fig.~\ref{fig:motivation}  the tidal disruption of two stars by a massive black hole located at the coordinate 
origin (the simulation is discussed in more detail in \cite{rosswog22a}).
As the stars pass the black hole, they are ripped apart by the hole's tidal forces. One is only interested in the fate
of the gas from the disrupted stars, which initially covers only a minute fraction of $10^{-9}$ of the space shown in
Fig.~\ref{fig:motivation}. So, in a corresponding Eulerian simulation, one would need to waste essentially all of the
computational resources to simulate the uninteresting empty space. 
Another major advantage
of SPH is that it can be formulated in a way that Nature's conservation laws are enforced by construction
\cite{monaghan05,rosswog09b,springel10a,price12a,rosswog15c}. Having Nature's conservation laws
built into a simulation gives some confidence that the simulated system behaves similarly to what Nature does.
\\
Simulating gas flows with particles is still a relatively young field compared to Eulerian gas dynamics, and much
development is still ongoing, both in terms of methodology and in
terms of spreading into new research areas. For example, most
recently, SPH methods have found their way into general relativistic
hydrodynamics where the full spacetime is dynamically evolved together with the fluid, see \cite{rosswog21a,rosswog23a}, or Chap.~7 in \cite{bambi25}.  
SPH has initially been criticized for many issues, but essentially all of them
have seen major improvements achieved in recent times.  For example, SPH has often been criticized for being too dissipative. However, as derived 
from a Lagrangian, SPH contains zero dissipation, and the criticism goes back to the early days of SPH when simple artificial dissipation schemes were used, 
in which dissipation was always switched on, whether
needed or not. In recent years, much effort has been spent
to cure this problem, e.g., by using time-dependent dissipation parameters that only reach substantial values when needed,
but not otherwise \cite{morris97,rosswog00,cullen10,rosswog15b,rosswog20b} or by using slope-limited reconstruction in the
artificial dissipation terms \cite{frontiere17,rosswog20a,rosswog21a,rosswog23a,sandnes24}. The latter approach is very effective even if large constant 
dissipation parameters are used. For example, some weakly triggered Kelvin-Helmholtz instabilities do not grow when standard, constant dissipation 
parameters {\em without} reconstruction are used, but the same initial conditions lead to
a healthy growth of instability when reconstruction {\em is} applied; see Fig.~20 in \cite{rosswog20a}.\\
As an alternative to artificial dissipation, one can also
implement Riemann solvers into SPH \cite{inutsuka02,parshikov02,sirotkin13,puri14,meng21} to produce sufficient entropy in shocks. However, with 
this approach, one also needs to ensure that not too much unwanted dissipation is introduced. For example, simply treating particle pairs as Riemann 
problems, where the left and the right states are given by the particle properties, leads to very dissipative hydrodynamics unless other measures such 
as limiters or reconstruction techniques similar to Finite Volume schemes are applied. Often, Riemann solver approaches are only benchmarked against 
shocks, where they perform, by construction, very well. However, such approaches can still be way too dissipative to accurately model the growth of 
weakly triggered fluid instabilities.\\
It has also turned out that the still frequently used cubic spline kernel is not a great choice after all, but
substantially better kernels are readily available \cite{wendland95,cabezon08,dehnen12,rosswog15b} and replacing
the kernel function only requires very small changes to existing codes. It has further been realized that much more accurate gradient 
estimates \cite{garcia_senz12,cabezon12a,rosswog15b} than the standard kernel gradients can be obtained at a 
moderate additional cost and even without sacrificing the highly valued kernel
anti-symmetry, $\nabla_a W_{ab}= - \nabla_b W_{ab}$, which is crucial for good numerical conservation,  see e.g. Sec. 2.4 in \cite{rosswog09b}.\\
Several of the above improvements of SPH have borrowed techniques that are traditionally used in Finite Volume schemes, such as reconstruction, 
slope-limiting, or Riemann solvers. For example, recent work has explored the application of weighted essentially non-oscillatory (WENO) strategies to 
SPH; see e.g. \cite{antona21,avesani21,vergnaud23,antona24}.  Apart from "enhanced SPH", there is also a class of methods that tries to consistently 
formulate the ideal gas dynamics equations from the beginning as in Finite Volume methods, though for particles rather than for meshes, see e.g.  
\cite{benmoussa99,vila99,hietel00,junk03,gaburov11} for some pioneering work. This also comes with a broad variety of names for relatively similar methods, 
some of which are "proper SPH", some are "SPH with elements from Finite Volume methods", and yet others are proper Finite Volume discretizations of the 
conservation laws written for particles. Among these latter methods, there are formulations for stationary particles (Eulerian), for particles that move with the 
fluid velocity (Lagrangian) or for particles that move with any other velocity (Adaptive Lagrangian Eulerian or ALE methods); see, e.g. \cite{ramirez22}. Therefore, 
the boundaries between different methods are blurred, and it becomes a matter of semantics or taste  how to call a given method. In summary,  many of the 
issues that SPH has been criticized for can be considered as essentially solved, and modern SPH formulations can accurately solve very challenging gas dynamics 
problems\footnote{For the publication \cite{rosswog20a}, the author collected in the computational astrophysics community a large set of test problems
"that SPH cannot do", but please see the (excellent) results in this paper.}, for example \cite{frontiere17,rosswog20a,sandnes24}.\\
One issue that has been strongly improved by the above measures but is usually still not exactly enforced in the standard
SPH approaches is the lack of zero-order consistency. In simple words, the SPH approximation makes use of
weighted kernel sums over neighboring particles, but the weights in these sums do not add up exactly to
unity; see our discussion in Section ~\ref{subsec:RPK}. Whether this is an issue in practical applications or not depends on the kernels/neighbor numbers 
that are used and the tested problem considered, but it is desirable to avoid the issue in the first place. Without zeroth-order consistency, not even constant functions are reproduced exactly.\\
In this paper, we formulate and explore a new approach, in which we start from one of the commonly used SPH discretizations, but a) we introduce Roe's approximate solver, b) we identify the dissipative terms in the Riemann solver, and in these terms, we use slope-limited reconstruction, and finally, c) we use linearly reproducing kernels so that constant and linear functions are reproduced to machine precision. We carefully scrutinize our implementation of the reproducing kernels and put the new method to the test with many very challenging benchmark tests, ranging from shocks over fluid instabilities to vorticity creating  Schulz-Rinne shocks  \cite{schulzrinne93a}. 
All of these tests are performed in three spatial dimensions. In Sec.~\ref{sec:method} we discuss all the elements involved in our approach, and in Sec.~\ref{sec:results}, we present and discuss our benchmark tests before we summarize and conclude in Sec.~\ref{sec:summary}.

\section{Methodology}
\label{sec:method}

\subsection{Particle hydrodynamics formulation with a Riemann solver}
In the following, we will label the particles with $a,b$ and
$k$, where $a$ is usually the particle of interest, $b$ a neighbor particle, and $k$ can be either of them. We also use the Einstein sum convention with repeated indices, which implies a sum from 1 to 3, and we usually use $i,j,l$ and $m$ for the summation indices. Since we are working in a non-relativistic context and do not have to distinguish between co- and contravariant vectors, it has no particular meaning whether an index is written as a subscript or a superscript; readability considerations mostly guide this.

\label{subsec:Riemann_SPH}
Our starting point is a common  set of Smoothed Particle Hydrodynamics (SPH) equations
\cite{parshikov02,liu03,meng21}
\bea
\rho_a&=&   \sum_b m_b \bar{W}_{ab}\label{eq:dens_sum}\\
\frac{d\vec{v}_a}{dt}&=& - \sum_b m_b \frac{P_a + P_b}{\rho_a \rho_b} \nabla_a \bar{W}_{ab}\label{eq:dvdt1}\\
\frac{du_a}{dt}&=& \frac{1}{2} \sum_b m_b \frac{P_a + P_b}{\rho_a \rho_b} \vec{v}_{ab} \cdot \nabla_a \bar{W}_{ab}.
\label{eq:dudt1}
\eea
In Eqs.~(\ref{eq:dvdt1}) and (\ref{eq:dudt1}), the index $a$ at the nabla operator indicates that it is evaluated with respect
to particle $a$ (and not $b$).
As usual, one has the choice between either solving the continuity equation directly or obtaining the density $\rho_a$ through a kernel-weighted sum over neighbors, see Eq.~(\ref{eq:dens_sum}), which is our choice here.
The continuity equation approach has been found to be advantageous in some recent studies \cite{meng21,sandnes24},
but we choose here the summation approach, since it is {\em guaranteed} to deliver a strictly positive density due to the
positive definite kernels $W$ that we are using. Alternatives to this approach will be explored in the future.
The size of the kernel support is determined by the smoothing length $h$, $P$ denotes the gas pressure, $u$ is the specific internal energy, and $\vec{v}_{ab}= \vec{v}_a - \vec{v}_b$. Often, the density estimation, Eq.~(\ref{eq:dens_sum}), is performed "as locally as possible" in the sense that the kernel of a particle $a$ is evaluated with its smoothing length, $h_a$. To be consistent with our later choices, see Sec. \ref{subsec:RPK}, we choose here 
\be
\bar{W}_{ab}= \frac{W(|\ra - \rb|/h_a )+ W(|\ra - \rb|/h_b)}{2} \quad
{\rm and} \quad \nabla \bar{W}_{ab}= \frac{\nabla W(|\ra - \rb|/h_a )+ \nabla W(|\ra - \rb|/h_b)}{2}.
\label{eq:bar_Wab}
\ee
\\
While we have written the above equations in a commonly used way, it is worth keeping in mind that the kernel gradient can 
be explicitly written (for some smoothing length $h$) as
\be
\nabla_a W_{ab}(h)= \frac{\p W(q_{ab})}{\p q_{ab}} \frac{\p q_{ab}}{\p\vec{r}_a}= \frac{\p W(q_{ab})}{\p q_{ab}} \frac{\hat{e}_{ab}}{h},
\label{eq:std_gradW}
\ee
where $q_{ab}= |\vec{r}_a -\vec{r}_b|/h$ and $\hat{e}_{ab}= (\vec{r}_a -\vec{r}_b)/|\vec{r}_a -\vec{r}_b|$ is the unit vector pointing from particle $b$
to particle $a$. With this, Eq.(\ref{eq:dudt1}) can also be written as
\be
\frac{du_a}{dt}= \frac{1}{2} \sum_b m_b \frac{P_a + P_b}{\rho_a \rho_b} \left(\vec{v}_a \cdot \hat{e}_{ab} - \vec{v}_b \cdot \hat{e}_{ab}  \right) \frac{1}{h_{ab}} \frac{\p W}{\p q_{ab}}
=\frac{1}{2} \sum_b m_b   \frac{P_a + P_b}{h_{ab} \; \rho_a \rho_b}  \frac{\p W}{\p q_{ab}} \left(\tilde{v}_a - \tilde{v}_b \right),
\ee
where the velocities projected on the connection line are marked with
a tilde.\\
In gas dynamics, seemingly harmless sound waves can steepen into
shock waves, and for their correct treatment, dissipation is needed.
So far, however, the equations are entirely nondissipative, and to handle shocks, they need to be enhanced by mechanisms that produce sufficient (but not too much) 
dissipation to obtain non-oscillatory solutions in shocks. As outlined in the introduction, an often used approach is artificial viscosity, which has seen many improvements 
in recent times, including more accurate ways to steer dissipation parameters \cite{cullen10,rosswog15b,rosswog20b}, applying reconstruction techniques in dissipative 
terms \cite{frontiere17,rosswog20a,sandnes24}. These new developments have also been transferred to (general) relativistic hydrodynamics \cite{rosswog21a,rosswog22b,rosswog23a}.
These new approaches have turned out to be robust and accurate and to produce very little unwanted dissipation. 
In our understanding, such modern artificial dissipation approaches are on par with approximate Riemann solvers. \\
\begin{figure} 
   \centering
   \centerline{
   \includegraphics[width=\columnwidth]{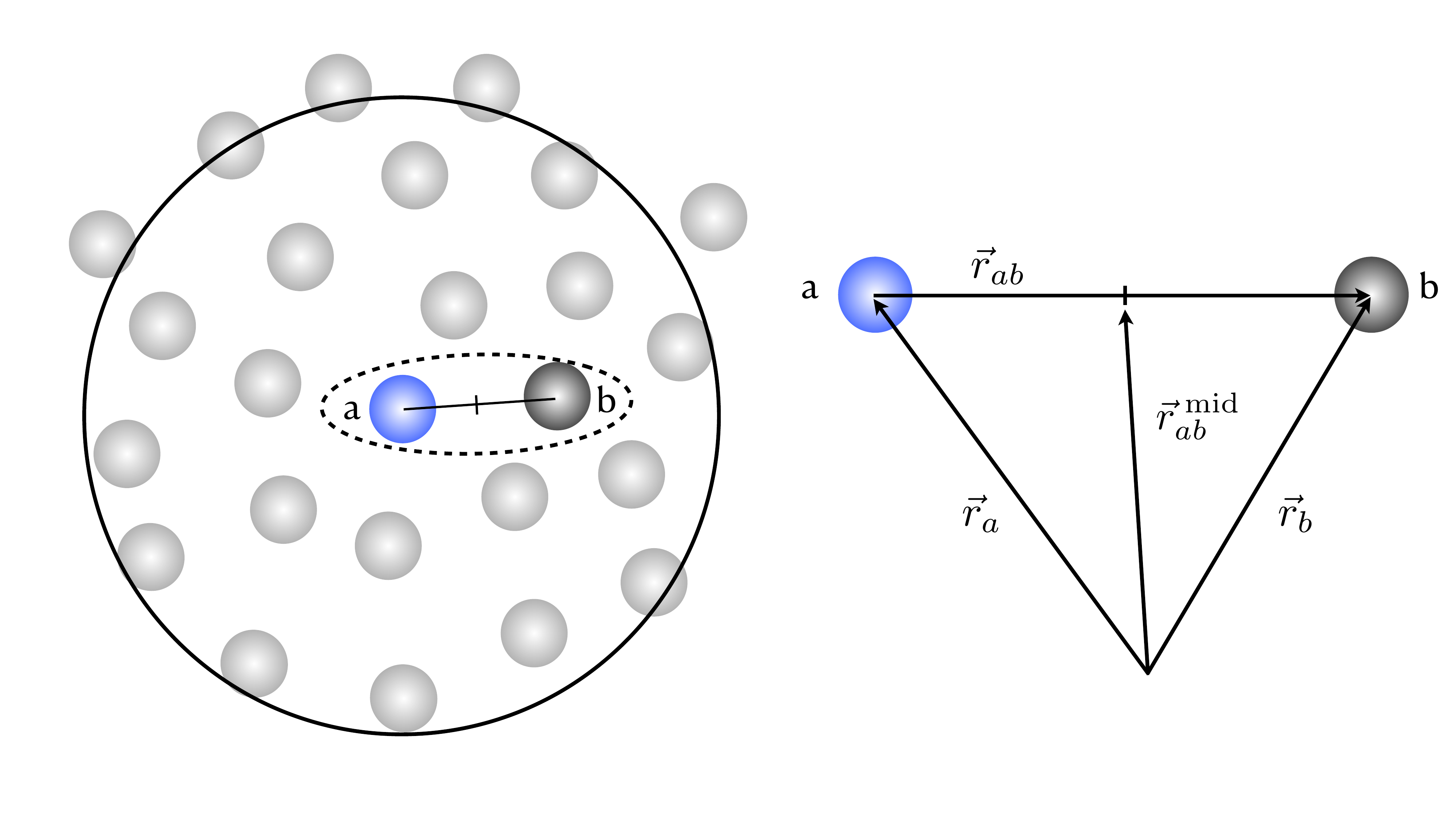}}
    \caption{To add dissipation,  Riemann problems are solved for each particle and its neighbour particles.}
   \label{fig:Riemann_sketch}
\end{figure}
Despite the very good performance of modern artificial viscosity schemes, we want to explore here the case where dissipation is provided by a Riemann solver \cite{toro09}.
The main idea is to solve (approximately) a one-dimensional Riemann problem at the midpoint between each pair of
interacting particles $a$ and $b$, as sketched in Fig.~\ref{fig:Riemann_sketch}.
More specifically, we  replace averages of particle values by the solution of a Riemann problem
\be
\frac{P_a + P_b}{2} \rightarrow P_{ab}^\ast \quad {\rm and} \quad  \frac{\tilde{v}_a + \tilde{v}_b}{2} \rightarrow v_{ab}^\ast, 
\label{eq:R_subs}
\ee
where the $\ast$ labels the contact discontinuity state in a Riemann
problem,  see e.g. \cite{toro09}, and  the $ab$-index refers to the
solution between state $a$ and $b$. 
As before, the tilde denotes the velocities projected onto the line connecting two particles
\be
\tilde{v}_a= \vec{v}_a \cdot \hat{e}_{ab} \quad {\rm and} \quad  \tilde{v}_b= \vec{v}_b \cdot \hat{e}_{ab}.
\ee
With the substitutions Eq.~(\ref{eq:R_subs}) we have
\be
P_a + P_b = 2 P_{ab}^\ast \quad {\rm and} \quad \tilde{v}_a - \tilde{v}_b= 2 (\tilde{v}_a - v_{ab}^\ast)
\ee
so that the hydrodynamics equations, now {\em with} dissipation, read
\bea 
\rho_a&=&   \sum_b m_b \bar{W}_{ab}\label{eq:dens_sum2}\\
\frac{d \vec{v}_a}{dt} &=& - \frac{2}{\rho_a} \sum_b V_b P_{ab}^\ast \nabla_a \bar{W}_{ab}\label{eq:dvdt2}\\
\frac{d u_a}{dt}&=&  \frac{2}{\rho_a}  \sum_b  V_b  P_{ab}^\ast  (\vec{v}_a - \vec{v}^\ast_{ab}) \cdot \nabla_a \bar{W}_{ab}\label{eq:dudt2},
\eea
where we have used $V_b= m_b/\rho_b$ and $\vec{v}^\ast_{ab}= v_{ab}^\ast \hat{e}_{ab}$.

\subsection{Dissipation in the Roe solver}
\label{subsec:diss_Roe}
We use Roe's approximate Riemann solver \cite{roe86} for the star state:
\bea
v_{ab}^\ast &=& \frac{1}{2} \left( (\vec{v}_a + \vec{v}_b) \cdot \hat{e}_{ab} + \frac{P_b - P_a}{C_{\rm RL}}\right)\label{eq:vstar_Roe}\\
P_{ab}^\ast &=& \frac{1}{2} \left( P_a + P_b  +  C_{\rm RL} \frac{ (\vec{v}_b - \vec{v}_a) \cdot \hat{e}_{ab}} {C_{\rm RL}} \right) \label{eq:Pstar_Roe},
\eea
where the "densitized" Roe-averaged Lagrangian sound speed (the dimension is density times velocity) is
\be
C_{\rm RL}= \frac{c_{s,a} \rho_a \sqrt{\rho_a} + c_{s,b} \rho_b \sqrt{\rho_b}}{\sqrt{\rho_a} + \sqrt{\rho_b}}
\ee
 and $c_{\rm s,k}$  is the sound speed of particle $k$.
If, for a moment, we ignore the terms that involve the pressure and velocity differences in Eqs.~(\ref{eq:vstar_Roe}) and (\ref{eq:Pstar_Roe}), we have
\be
v_{ab}^\ast \approx \frac{\vec{v}_a + \vec{v}_b}{2} \cdot \hat{e}_{ab}  \quad {\rm and} \quad
P_{ab}^\ast \approx \frac{ P_a + P_b}{2} .
\ee
By inserting these expressions into Eqs.~(\ref{eq:dvdt2}) and (\ref{eq:dudt2}) we obviously recover the inviscid equations
(\ref{eq:dvdt1}) and (\ref{eq:dudt1}), therefore {\em the terms involving the differences in
Eqs.~(\ref{eq:vstar_Roe}) and (\ref{eq:Pstar_Roe}) are responsible for the dissipation.} 
One may thus try to cure potentially excessive dissipation by modifying these terms. We approach this issue here
by applying the reconstructed pressure and velocity values {\em in the dissipative terms}. So, for perfectly reconstructed smooth flows, where the reconstructed values on both sides of the midpoint are the same and, therefore, the dissipative terms vanish, one effectively solves the inviscid hydrodynamics equations.
  
\subsection{Reconstruction in the dissipative terms}
\label{subsec:reconst}
Our strategy to reduce dissipation is to use, in the dissipative terms of Eqs.~(\ref{eq:vstar_Roe}) and (\ref{eq:Pstar_Roe}),
values of $P$ and $\vec{v}$ that are reconstructed to the midpoint of each particle pair,  $\vec{r}_{ab}^{\rm mid}=
0.5(\vec{r}_a + \vec{r}_b)$, see Fig.~\ref{fig:Riemann_sketch}, left panel.
So, at each interparticle midpoint, one has $P$/$\vec{v}$ values that are reconstructed once from the $a$-side and once from the $b$-side. Explicitly, the linearly reconstructed 
velocity components read
\bea
v_a^{i, \rm rec} &=& v^i_a - \frac{1}{2} \Psi(\nabla v^i_a,\nabla v^i_b) \cdot \vec{r}_{ab}\label{eq:rec1}\\
v_b^{i, \rm rec} &=& v^i_b + \frac{1}{2} \Psi(\nabla v^i_a,\nabla v^i_b) \cdot \vec{r}_{ab},
\eea 
the specific internal energy is
\bea
u_a^{\rm rec} &=& u_a - \frac{1}{2} \Psi(\nabla u_a,\nabla u_b) \cdot \vec{r}_{ab}\label{eq:rec3}\\
u_b^{\rm rec} &=& u_b + \frac{1}{2} \Psi(\nabla u_a,\nabla u_b) \cdot \vec{r}_{ab}
\eea
and the density
\bea
\rho_a^{\rm rec} &=& \rho_a - \frac{1}{2} \Psi(\nabla \rho_a,\nabla \rho_b) \cdot \vec{r}_{ab}\\
\rho_b^{\rm rec} &=& \rho_b + \frac{1}{2} \Psi(\nabla \rho_a,\nabla \rho_b) \cdot \vec{r}_{ab} \label{eq:rec6}.
\eea
With the reconstructed values of internal energy and density, we can calculate the
corresponding pressure values via our equation of state, $P= (\gamma-1)\rho u$.\\
The quantity $\Psi$ in the above reconstruction equations is a suitable slope limiter function. Commonly used slope limiters are  \texttt{minmod} 
\be
\Psi_{\rm mm}(x,y)= \frac{1}{2}\left[\rm{SGN}(x) + \rm{SGN}(y)\right] \; {\rm MIN}(|x|,|y|),
\ee
the \texttt{vanLeer limiter} \cite{vanLeer77}
\be
\Psi_{\rm vL}(x,y)= \begin{cases}\frac{2 x y}{x+ y} \quad & {\rm if } \; x y > 0\\
                                               0 & {\rm otherwise},
                         \end{cases}                      
\ee
the vanLeer monotonized Central (\texttt{vanLeerMC}) \cite{vanLeer77}
\be
\Psi_{\rm vLMC}(x,y)= \begin{cases} \rm{SGN}(x) \; \rm{MIN}\left[\frac{|x+y|}{2}, 2 |x|, 2 |y|\right] \quad & {\rm if } \; x y > 0 \\
                                                      0 & {\rm otherwise},
                              \end{cases}                        
\ee
and the \texttt{vanAlbada limiter} \cite{vanAlbada82}
\be
\Psi_{\rm vA}(x,y)= \begin{cases} \frac{(x^2 + \epsilon^2) y + (y^2 + \epsilon^2) x}{x^2 + y^2 + 2 \epsilon^2} \quad & {\rm if } \; x y > 0,\\
                                                 0 & {\rm otherwise}.
                          \end{cases}                        
\ee
The latter is insensitive to the exact value of $\epsilon$, here, we
use $\epsilon^2= 10^{-6}$.  In Eqs.~(\ref{eq:rec3}) to
  (\ref{eq:rec6}) the slope limiter $\Psi$ is to be applied to
  each component of the gradients.
%
\begin{figure}
   \centering
   \centerline{
   \includegraphics[width=8cm]{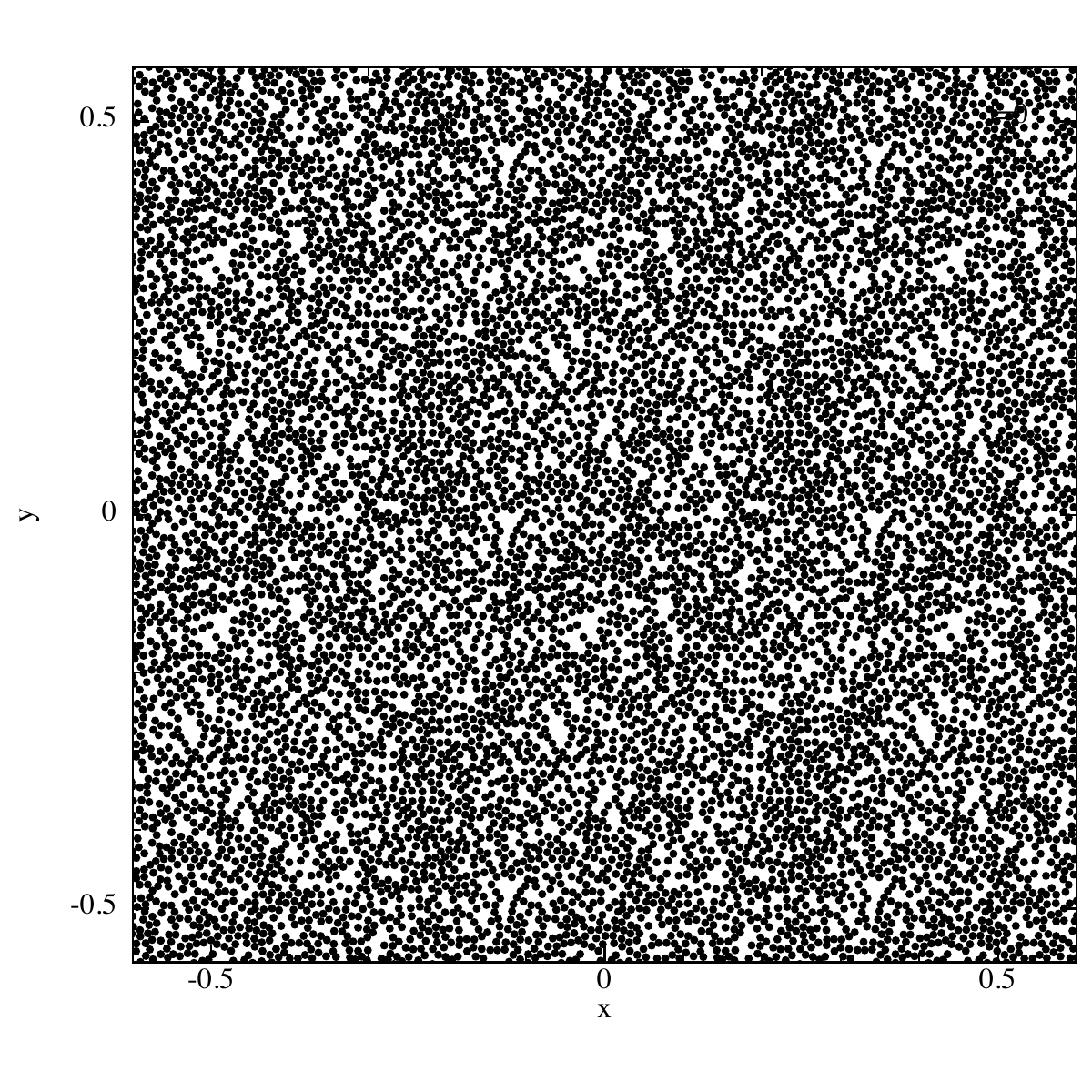} }
    \caption{The particle distribution that is used to scrutinize the linear reproducing kernel method. The 3D particle distribution has been set up using a Centroidal Voronoi Tessellation; shown is a slice with a thickness of $|z| < 6 \times 10^{-3}$.}
   \label{fig:part_positions}
\end{figure}
%
\begin{figure*} 
   \centering
   \centerline{
   \includegraphics[width=8cm]{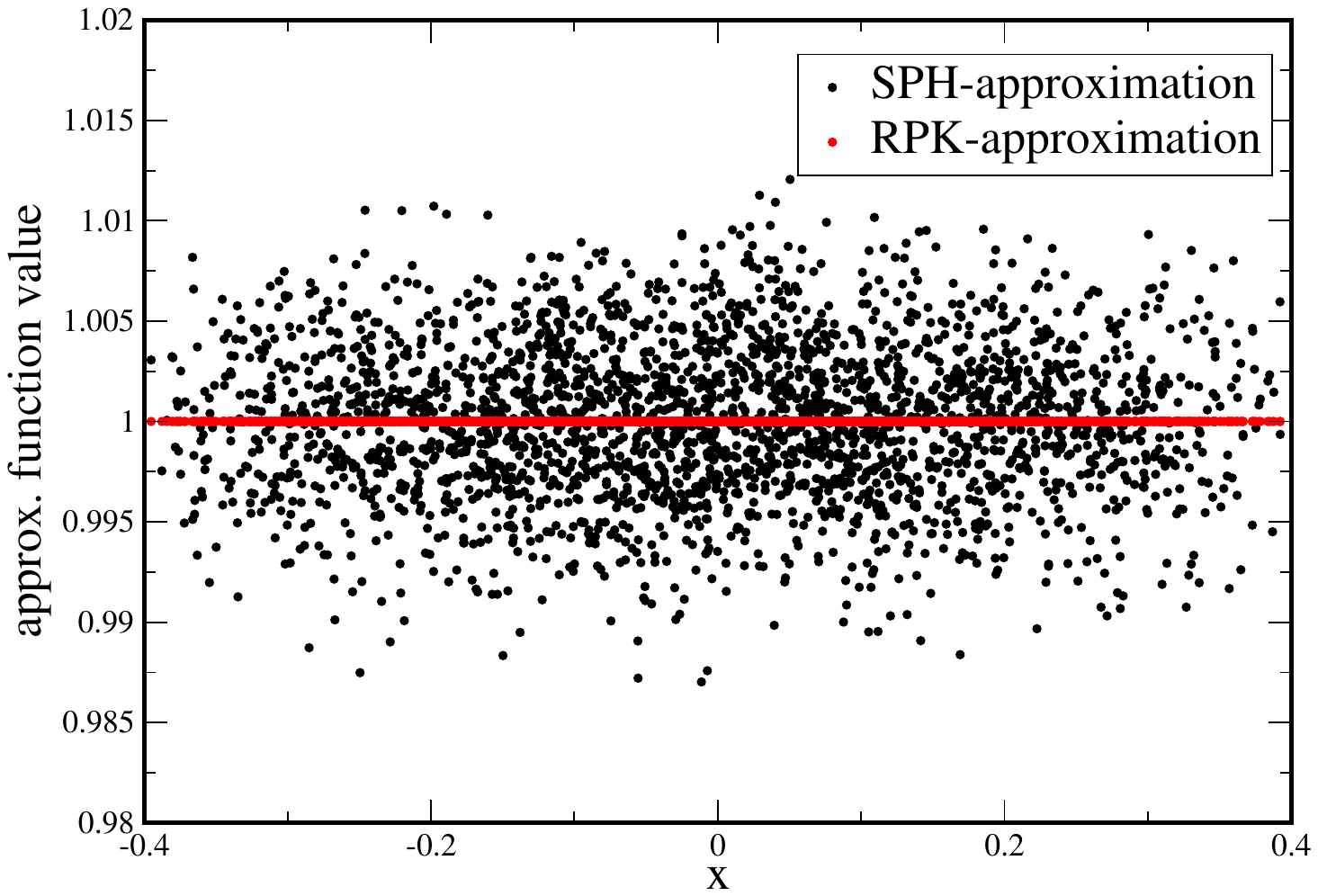}  \hspace*{-0.6cm}
   \includegraphics[width=8cm]{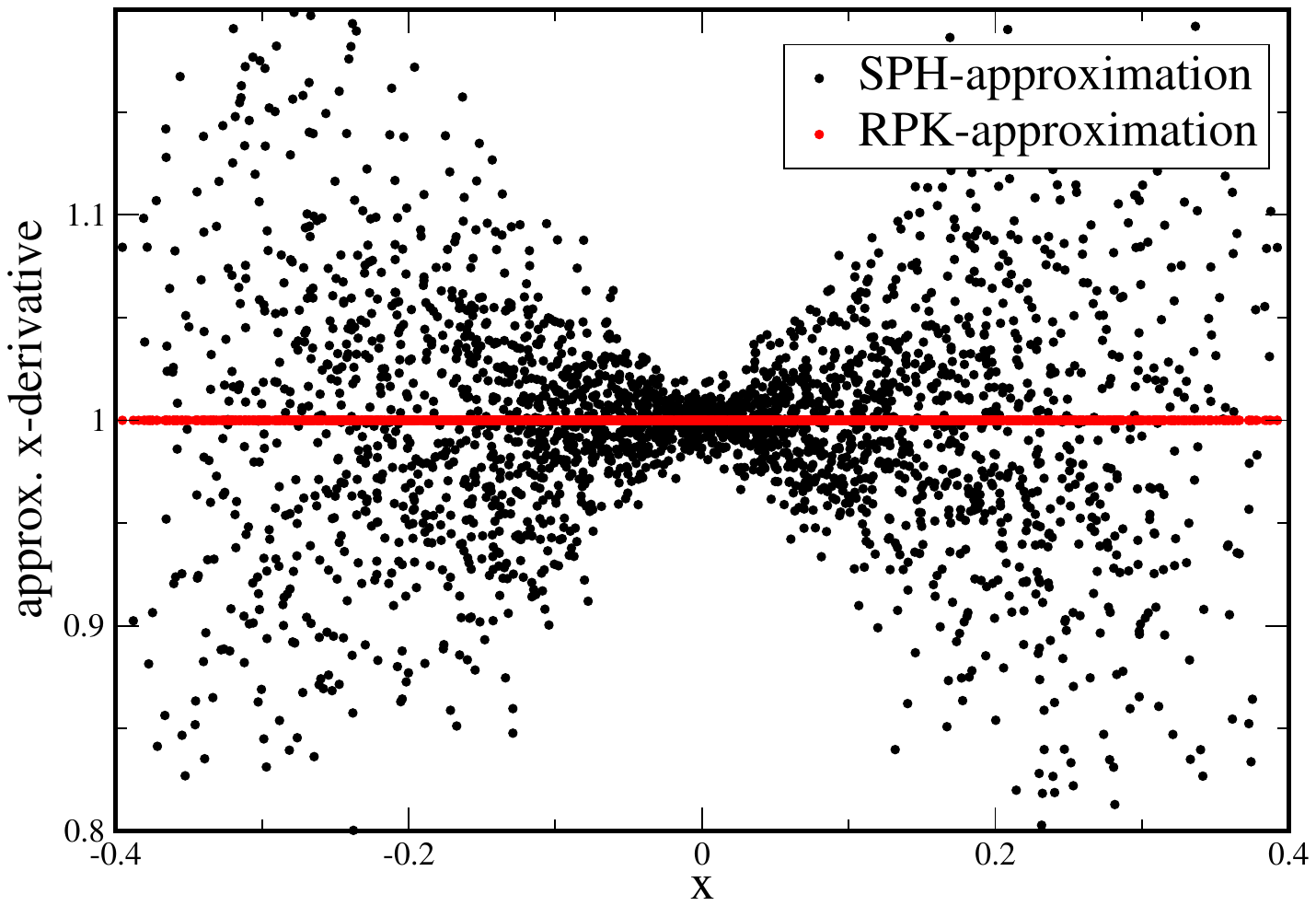} }
    \caption{Approximation of function value (left) and derivative in $x$-direction (right), the exact result is in both cases $=1$.
    Each time we show the SPH-approximation in black, the 
      reproducing kernel (RPK-) result in red.}
   \label{fig:func_approx}
\end{figure*}

\subsection{Linearly reproducing kernel (RPK) approximations}
\label{subsec:RPK}
One of the main criticisms of the standard SPH method is that it, in general, does not exactly reproduce constant or linear functions. 
In the standard SPH-discretization the approximation of a function $f(\vec{r})$ reads
\bea
\tilde{f}(\vec{r})= \sum_b \; V_b \; f_b \; W(\vec{r} - \vec{r}_b,h(\vec{r}))= \sum_b f_b \; \Phi_b(\vec{r})
\label{eq:discrete_SPH_approx},
\eea
where we have abbreviated $m_b/\rho_b$ as $V_b$ and the smoothing length $h$
should be thought of as a function of $\vec{r}$. After the second equal sign, we have abbreviated
the product $V_b W(\vec{r} - \vec{r}_b,h(\vec{r}))$ as weight function $\Phi_b(\vec{r})$.  As an example,
if all function values would be the same, $f_b= f_0$, then the function approximation should yield $f_0$,
but the standard SPH approximation finds
\be
\tilde{f}(\vec{r})= f_0 \sum_b \frac{m_b}{\rho_b}  W(\vec{r} - \vec{r}_b,h) \approx f_0,
\ee
which is approximately, but not exactly, equal to the desired value, because the sum is not guaranteed
to yield exactly unity. More systematically,  one can expand $f_b$ (i.e. the function $f$ at the position $b$) 
around $\vec{r}$, insert this into Eq.~(\ref{eq:discrete_SPH_approx}) to find
\bea
\tilde{f}(\vec{r})= \sum_b \left[ f(\vec{r}) + (\nabla f)(\vec{r}) \cdot (\vec{r_b} - \vec{r}) + {\rm h.o.t.} \right]\; \Phi_b(\vec{r})
                        = f(\vec{r}) \sum_b \Phi_b(\vec{r}) + (\nabla f)(\vec{r}) \cdot \sum_b (\vec{r_b} - \vec{r}) \Phi_b(\vec{r}) + {\rm h.o.t.} ,
 \eea
where higher order terms are abbreviated as ``h.o.t''.
This implies, not too surprisingly, that for a good approximation, $\tilde{f}(\vec{r})\approx f(\vec{r})$, the lowest order
{\em consistency relations}
\be
\sum_b \Phi_b(\vec{r})= 1 \quad {\rm and} \quad \sum_b (\vec{r}_b - \vec{r})^i \;  \Phi_b(\vec{r}) = 0, \label{eq:consistency}
\ee
should be fulfilled. Standard SPH usually fulfils this well in
initial conditions  (where particles are often placed on some type of lattice), but it does not enforce these
conditions during evolution. The conditions are well fulfilled when good kernels (e.g., of the Wendland
family \cite{wendland95}) with large neighbour numbers are used \cite{rosswog15b}, but this, of course, comes at the price
of expensive sums over many neighboring particles.\\
\begin{figure*}
   \centering
   \centerline{
    \includegraphics[width=8cm]{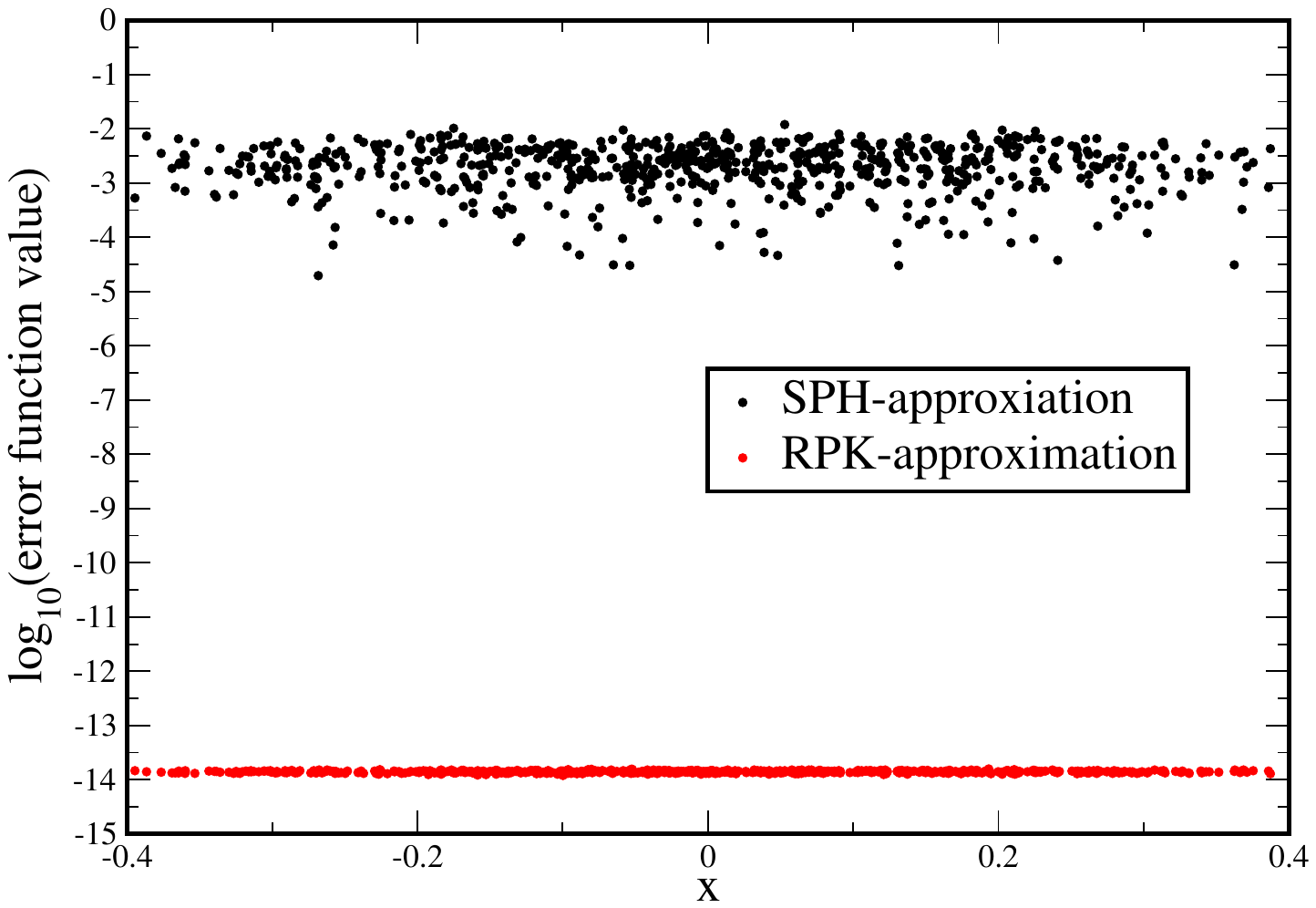}  \hspace*{-0.6cm}
    \includegraphics[width=8cm]{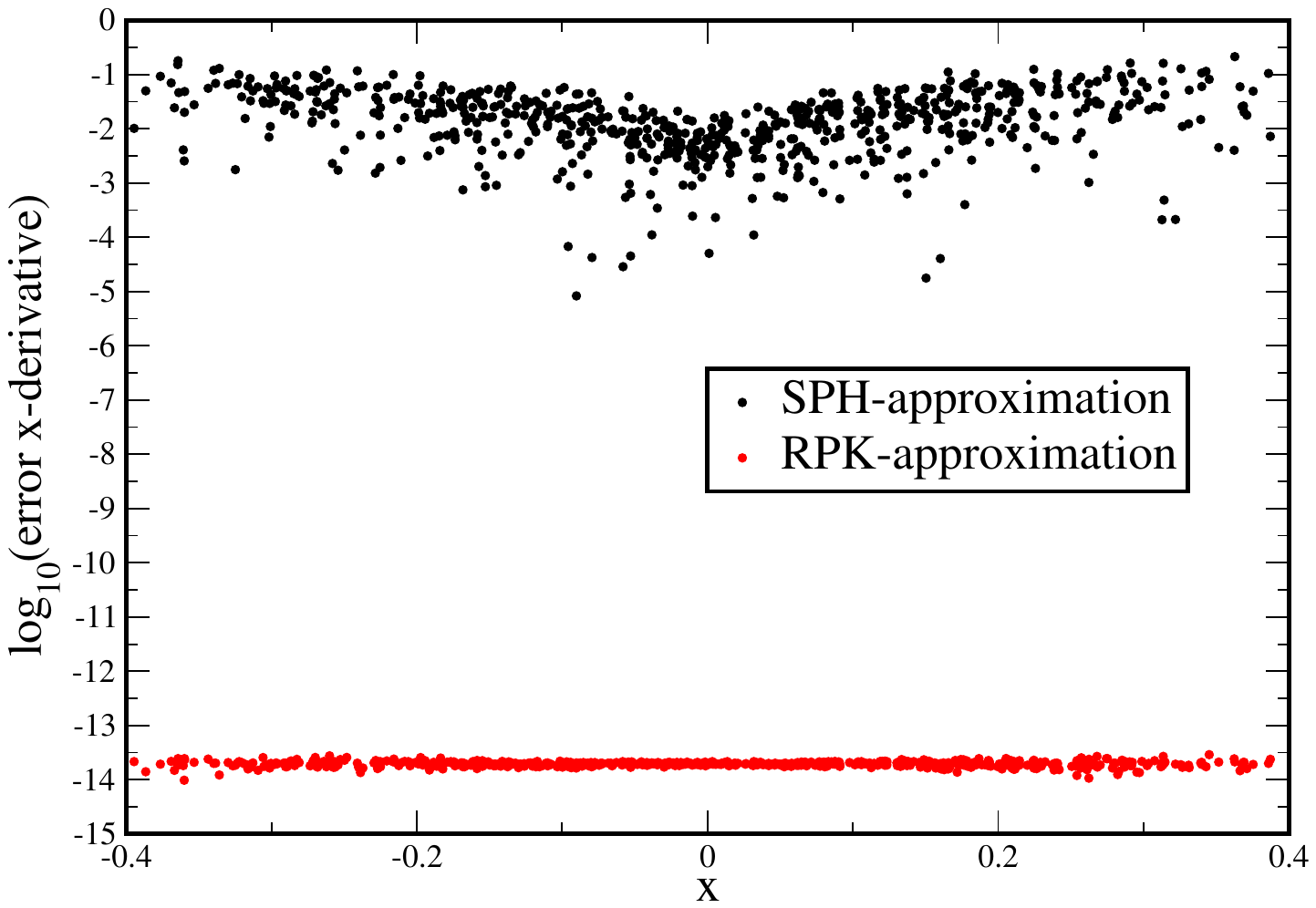}}
    \caption{Logarithm of the function error (left) and of the error in the $x$-derivative approximation (right).
    Each time we show the SPH-approximation in black and the 
      reproducing kernel (RPK-) result in red.}
   \label{fig:SPH_RPK_errors}
\end{figure*}
The consistency relations Eqs.~(\ref{eq:consistency}), however,  can also be enforced by construction, e.g. in the so-called reproducing 
kernel method \cite{liu95}. One can, for example, enhance the 
kernel functions with additional parameters $A$ and $B^i$,
\be
\mathcal{W}_{ab}(\vec{r}_{ab})= A_a \left[1 + B_a^i \; (\vec{r}_{ab})^i\right] \bar{W}_{ab},
\label{eq:RPK_kernel}
\ee 
where $\bar{W}_{ab}$ is given by Eq.~(\ref{eq:bar_Wab}). One can then determine the four unknown numbers $A$ and $B^i$
by enforcing the four discrete consistency relations so that 
\be
\sum_b V_b \; \mathcal{W}_{ab} = 1 \quad {\rm and} \quad
\sum_b V_b \; (\vec{r}_{ab})^i \; \mathcal{W}_{ab} = 0
\label{eq:consistency_relations}
\ee
is fulfilled at every particle position $\vec{r}_a$. Since these kernels $\mathcal{W}_{ab}$ reproduce by 
construction constant and linear functions exactly (i.e. to floating point accuracy), they are usually referred 
to as (linearly) reproducing kernels (RPKs). Reconstructing kernels for higher-order polynomials could be 
designed similarly, but at the price of lengthy and computationally expensive-to-evaluate expressions.
We, therefore, restrict ourselves here to linear order.
Since the $B^i$ are in general non-zero, the kernels
$\mathcal{W}_{ab}$  are not guaranteed to be
radial as standard SPH kernels, and therefore, angular momentum conservation is difficult to enforce exactly. 
In SPH, angular momentum conservation is usually a consequence of the interparticle forces pointing along the connecting vectors between a particle
$a$ and a particle $b$, $\hat{e}_{ab}$, together with the kernel being anti-symmetric, $\nabla_a W_{ab}= - \nabla_b W_{ab}$,
see Sec.~2.4 in \cite{rosswog09b} for a detailed discussion.
It is, however, still possible to write a set of equations that conserves energy and momentum
(and mass if density summation is used), but not necessarily angular momentum. In practice, however, this does not seem
to be a major concern; for example, the authors of \cite{frontiere17}  find with their artificial viscosity-based RPK-SPH approach in typical tests violations 
of exact angular momentum conservation on the sub-percent level.\\
The gradient of $\mathcal{W}_{ab}$ can be calculated in a
straight-forward way as 
\be
(\nabla_a)^k  \mathcal{W}_{ab} =  A_a \; B_a^k \; \bar{W}_{ab} + A_a \left(1+ B_a^i (\vec{r}_{ab})^i\right) \nabla_a^k \bar{W}_{ab} +
                                                                                                  \left( 1+ B_a^i (\vec{r}_{ab})^i \right) \bar{W}_{ab} (\p_k A)_a + A_a \; (\vec{r}_{ab})^i \; (\p_k B)_a^i  \; \bar{W}_{ab}
\label{eq:nabla_RPK_a}
\ee
Now, taking the nabla operator concerning particle $b$, one finds
\bea
(\nabla_b)^k  \mathcal{W}_{ba} =  &&   A_b \; B_b^k \; \bar{W}_{ba} + A_b \left(1+ B_b^i (\vec{r}_{ba})^i\right) \nabla_b^k \bar{W}_{ba} + \nonumber \\
                                                      && \left( 1+ B_b^i (\vec{r}_{ba})^i \right) \bar{W}_{ba} (\p_k A)_b + A_b \; (\vec{r}_{ba})^i \; (\p_k B)_b^i  \; \bar{W}_{ba}\nonumber\\
                                                 =   &&  A_b \; B_b^k \; \bar{W}_{ab} + A_b \left(1+ B_b^i (\vec{r}_{ab})^i\right) \nabla_a^k \bar{W}_{ab} + \nonumber \\
                                                      &&   \left( 1- B_b^i (\vec{r}_{ab})^i \right) \bar{W}_{ab} (\p_k A)_b - A_b \; (\vec{r}_{ab})^i \; (\p_k B)_b^i  \; \bar{W}_{ab} ,                                                                                                    
\label{eq:nabla_RPK_b}
\eea
where we have used $\bar{W}_{ab}= \bar{W}_{ba}$, $\vec{r}_{ba}= -\vec{r}_{ab}$ and $\nabla_b^k \bar{W}_{ba}= - \nabla_a^k \bar{W}_{ab}$.
The parameters $A$ and $B^i$ and their derivatives, which are needed for $\p_k \mathcal{W}_{ab}$, can then be calculated
by straightforward algebra. We first define discrete moments  at position $\ra$ as
\bea
\left(M_0\right)_a &\equiv& \sum_b V_b \; \bar{W}_{ab} \\
\left(M_1^i\right)_a &\equiv& \sum_b V_b \; (\vec{r}_{ab})^i \bar{W}_{ab} \\
\left(M_2^{ij}\right)_a &\equiv& \sum_b V_b \; (\vec{r}_{ab})^i  \; (\vec{r}_{ab})^j  \; \bar{W}_{ab}
\eea
and their derivatives read
\bea
\left(\p_k M_0\right)_a &\equiv&  \sum_b V_b \; \nabla_a^k  \bar{W}_{ab} \\
\left(\p_k M_1^i\right)_a &\equiv& \sum_b V_b \left[ (\vec{r}_{ab})^i (\nabla_a^k \bar{W}_{ab}) + \delta^{ki} \bar{W}_{ab} \right]\\
\left(\p_k M_2^{ij}\right)_a &\equiv& \sum_b V_b \; [ (\vec{r}_{ab})^i  \; (\vec{r}_{ab})^j  \; (\nabla_a^k \bar{W}_{ab}) + 
(\vec{r}_{ab})^i \; \delta^{jk} \; \bar{W}_{ab} +   (\vec{r}_{ab})^j \; \delta^{ik} \; \bar{W}_{ab} ].
\eea
With the moments and their derivatives at hand, one can calculate the
kernel parameters 
\bea
A_a &=& \frac{1}{\left(M_0\right)_a - (M_2^{ij})^{-1}_a \; \left(M_1^i\right)_a \; (M_1^j)_a}\\
B_a^i&=& - (M_2^{ij})_a^{-1} \; (M_1^j)_a
\eea
and their somewhat lengthy but  otherwise straightforward calculable derivatives
\bea
\p_k A_a = -A^2_a [ (\p_k M_0)_a - 2 (M_2^{ij})_a^{-1}   (M_1^j)_a  \left( \p_k M_1^i \right)_a  + (M_2^{il})_a^{-1}   (\p_k M_2^{lm})_a (M_2^{mj})_a^{-1} (M_1^j)_a (M_1^i)_a ]
\eea
and
\be
\p_k B_a^i= -(M_2^{ij})_a^{-1}  \; (\p_k M_1^j)_a + (M_2^{il})_a^{-1} \; (\p_k M_2^{lm})_a  (M_2^{mj})_a^{-1} (M_1^j)_a. 
\ee
With the linearly reproducing kernels  now at hand, we can approximate a function $F$ via
\be
\tilde{F}(\ra)= \sum_b V_b \; F_b \; \mathcal{W}_{ab}
\ee
and its derivative via
\be
\p_k \tilde{F}(\ra)= \sum_b V_b \; F_b \; \p_k \mathcal{W}_{ab}.
\ee
While these expressions look very similar to standard SPH approximations, they 
exactly reproduce linear functions on a discrete level, which the SPH equations do not.

\subsubsection{Scrutinizing our implementation}
\begin{table}
\caption{Average errors in  the function and derivative
    approximation for the particle
  distribution shown in Fig.~\ref{fig:part_positions} for both the Smoothed Particle Hydrodynamics (SPH) 
  and the reproducing kernel (RPK) approach.}
\label{tab:error}       
\centering
\begin{tabular}{m{15ex}m{15ex}m{15ex}}
\hline\noalign{\medskip}
 to approximate   & $f=1$   &  $\partial_x f= 1$ \\\hline\hline\noalign{\medskip}
SPH (glass) & $3.1 \times 10^{-3}$ &  $4.3 \times 10^{-2}$\\\hline\noalign{\medskip}
RPK (glass) & $2.2 \times 10^{-14}$ & $1.9 \times 10^{-14}$ \\\hline\noalign{\medskip}
\end{tabular}
\end{table}
To test our implementation of the linear reproducing kernels, we start from a relatively regular, but not exactly uniform
particle distribution, sometimes referred to as "a glass". Our initial particle configuration, see Fig.~\ref{fig:part_positions},
has been produced via  a Centroidal Voronoi Tessellation (CVT; \cite{du05}) in the computational 
volume [0.5,0.5] $\times$ [0.5,0.5] $\times$ [0.5,0.5], and it is further regularized 
by additional sweeps according to the "artificial pressure method"
(APM), as described in the \Ma paper \cite{rosswog20a}, see especially Eq.~(40).
We perform the regularization sweeps for the inner regions while the outer regions remain as boundary particles on the original CVT setup.
The resulting particle distribution is therefore most regular in the centre, approaching the somewhat rougher original distribution as one moves
away from the centre. This is hard  to see by eye, but it is reflected in the SPH errors, as seen below.
We now assign functions $f$ to each particle, a) once a constant value $f^{(1)}=1$ and b) the other time we assign
a linear function that increases in the $x$-direction $f^{(2)}= x$ with a slope of unity.\\
Since by construction the consistency relations, Eq.~(\ref{eq:consistency_relations}), are fulfilled {\em at the particle positions} 
(i.e., not everywhere in space), we select every 100th particle in the inner regions (so that the absolute value of each coordinate is $<0.4$) 
and at these particles, labelled $a$, we calculate the standard SPH approximations
\be
f^{\rm SPH}_a= \sum_b V_b f_b \bar{W}_{ab} \quad {\rm and} \quad \left(\nabla f\right)^{\rm SPH}_a= \sum_b V_b f_b \nabla \bar{W}_{ab}
\ee
as well as the reproducing kernel approximations
\be
f^{\rm RPK}_a= \sum_b V_b f_b \mathcal{W}_{ab} \quad {\rm and} \quad \left(\nabla f\right)^{\rm RPK}_a= \sum_b V_b f_b \nabla \mathcal{W}_{ab}
\ee
and compare against the theoretical results of unity for both the function value and the derivative in $x$-direction.\\
For the  kernel $W$ that enters Eq.~(\ref{eq:bar_Wab}), we choose, here and in the rest of the paper,  a member of the 
family of harmonic-like kernels \cite{cabezon08}
\begin{align}
            W (x) = W^{\rm H}_8(x)= \frac{\sigma_8}{h^3} \left\{\begin{array}{llr}
        1, & \; x=0,\\[4mm]
        \left( \frac{\sin\left[ \frac{\pi}{2} x \right]}{\frac{\pi}{2} x}\right)^8
          & \; 0< x <2,\\[4mm]
        0, & \; {\rm else},
       \end{array}\right.
       \label{eq:WH8_kernel}
\end{align}
where $\sigma_8= 1.17851074088357$ in three dimensions. This kernel was chosen after performing a numerical test in which the ability to reproduce a known density was measured \cite{rosswog23a}. Our smoothing length is chosen at every time step so that we have exactly 220 contributing neighbor particles, which is a good compromise between accuracy and computational efficiency, see Fig.~1 in \cite{rosswog23a}.  We use a fast tree method \cite{gafton11} to assign the smoothing  lengths; for more details see \cite{rosswog20a}. The approximate values of the constant function (exact result =1; left panel) and the derivative (exact result = 1; right panel) are shown in Fig.~\ref{fig:func_approx} with the SPH approximation marked with black and the RPK approximation marked with red dots. The corresponding plot for the related errors is shown in Fig.~\ref{fig:SPH_RPK_errors}.
We find an average error in the function approximation of $3.1 \times
10^{-3}$ for the SPH and $2.2 \times 10^{-14}$ for the RPK case; see Tab.~\ref{tab:error}.
As a reference, we note that if the same test is performed for particles placed on a cubic lattice, we find an average error for the standard SPH approach of $7.3 \times 10^{-6}$.
On the "glass distribution" we find for the $x$-derivative an average error $|(\partial_x f)^{\rm SPH}- 1|=4.3 \times 10^{-2}$ in the SPH-case, better in the smoother central regions, worse further out,  and $1.9 \times 10^{-14}$ for RPK-case without noticeable dependence of the error on the location.
Again, for reference, the SPH derivative error on a cubic lattice is $1.1 \times 10^{-5}$, so much better than for the glass distribution. \\
In all fairness, it is worth stating that the SPH-approximation results for {\em this particle distribution} look rather poor, but the question is whether
such noisy particle distributions occur in an SPH simulation in the first place. Much recent work has been invested in designing simulation techniques so that
this {\em does not happen}, for example, by using Wendland kernels with the large neighbor number (typically several hundreds); see, e.g. \cite{rosswog15b,rosswog15c}.
This typically produces very regular particle distributions, where the error is much smaller than the ones shown here. However, the errors will never be smaller than what one finds for a regular lattice ($\sim 10^{-5}$). So, even for close-to-perfect particle distributions, the RPK approximations will be more than nine orders of magnitude more accurate than the standard SPH approach.

\begin{figure*}
  \hspace*{-0.7cm} \includegraphics[width=1.05\textwidth]{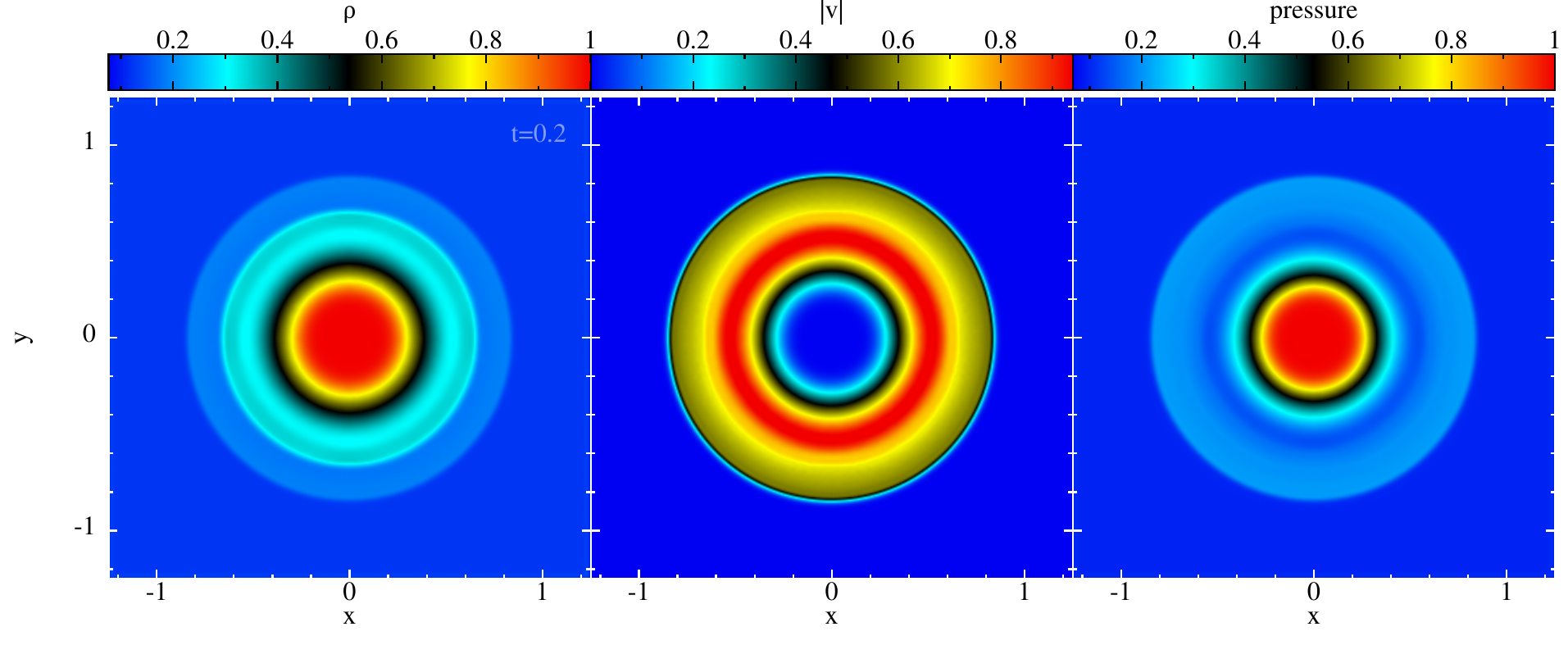} 
    \caption{Spherical blast wave problem 1: density, velocity and pressure, vanAlbada limiter. }
   \label{fig:Riemann1_colour}
\end{figure*}

\subsection{Final equation set}
\label{subsec:final_equations}
One desirable quantity for numerical conservation is the anti-symmetry of the kernel gradient
\be
\nabla_a W_{ab}(h_{ab})= - \nabla_b W_{ab}(h_{ab}),
\ee
which in typical SPH equations guarantees, in a straightforward way, the numerical conservation of energy and momentum, since in the time derivative of their total values, all terms cancel exactly, as seen, for example, in Sec. 2.4 in \cite{rosswog09b}. The usual standard kernel gradients point in the direction
of the line connecting two particles, see Eq.~(\ref{eq:std_gradW}); therefore, together with the kernel gradient anti-symmetry, the exact conservation of angular momentum can also be guaranteed. Due to the presence
of the vector $B^i$ in Eq.~(\ref{eq:RPK_kernel}) the kernels $\mathcal{W}$ are no longer guaranteed to
be radial, and therefore, angular momentum conservation cannot be guaranteed in the same way as in
standard SPH. The standard kernel gradient as it enters Eqs.~(\ref{eq:dvdt2}) and (\ref{eq:dudt2}) can also be written as
\be
\nabla_a \bar{W}_{ab}= \frac{\nabla_a W_{ab}(h_a) - \nabla_b W_{ba}(h_b)}{2},
\ee
where we have made use of the anti-symmetry of the kernel gradients. 
We now replace the gradients of the kernels $W$ by the
much more accurate gradients of the kernels $\mathcal{W}$, see Eqs.~(\ref{eq:nabla_RPK_a}) and (\ref{eq:nabla_RPK_b}),
or, in other words, we replace $\nabla_a \bar{W}_{ab}$ by
\be
(\nabla \mathcal{W})_{ab} \equiv \frac{1}{2}\left[ \nabla_a  \mathcal{W}_{ab} - \nabla_b  \mathcal{W}_{ba} \right],
\ee
so that our {\em final equation set} reads
\bea 
\rho_a&=&   \sum_b m_b \bar{W}_{ab}\label{eq:dens_sum3}\\
\frac{d \vec{v}_a}{dt} &=& - \frac{2}{\rho_a} \sum_b V_b P_{ab}^\ast  (\nabla \mathcal{W})_{ab} \label{eq:dvdt3}\\
\frac{d u_a}{dt}&=&  \frac{2}{\rho_a}  \sum_b  V_b  P_{ab}^\ast  (\vec{v}_a - \vec{v}^\ast_{ab}) \cdot (\nabla \mathcal{W})_{ab} \label{eq:dudt3}.
\eea
We also use the gradients $(\nabla \mathcal{W})_{ab}$ in the reconstruction process described in Sec.~\ref{subsec:reconst}.
As mentioned before, our equation set does not manifestly conserve angular momentum as standard SPH does, but our approach produces excellent results, at least for the broad set of challenging test cases that we show below. 
We integrate our evolution equations forward in time utilizing a  second order total variation diminishing Runge-Kutta scheme
  \cite{gottlieb98}. While this works very well in all tests, it is worth stating that time integration can, in principle, introduce numerical non-conservation. For detailed discussions of the role of time integration schemes in energy conservation and for compatible energy discretization schemes, we refer to the recent literature
  \cite{owen14,frontiere17,cercos_pita23}.

\section{Results}
\label{sec:results}
We will explore the performance of equations (\ref{eq:dens_sum3}) to (\ref{eq:dudt3}) in a set of challenging benchmarks involving shocks, instabilities, and complex Schulz-Rinne shocks with vorticity creation \cite{schulzrinne93a}. We will each time also explore the performance of the following slope limiters (in order decreasing dissipation): \texttt{minmod}, \texttt{vanAlbada}, and \texttt{vanLeerMC}.
\begin{figure*}
   \centering
   \centerline{
   \includegraphics[width=0.333\textwidth]{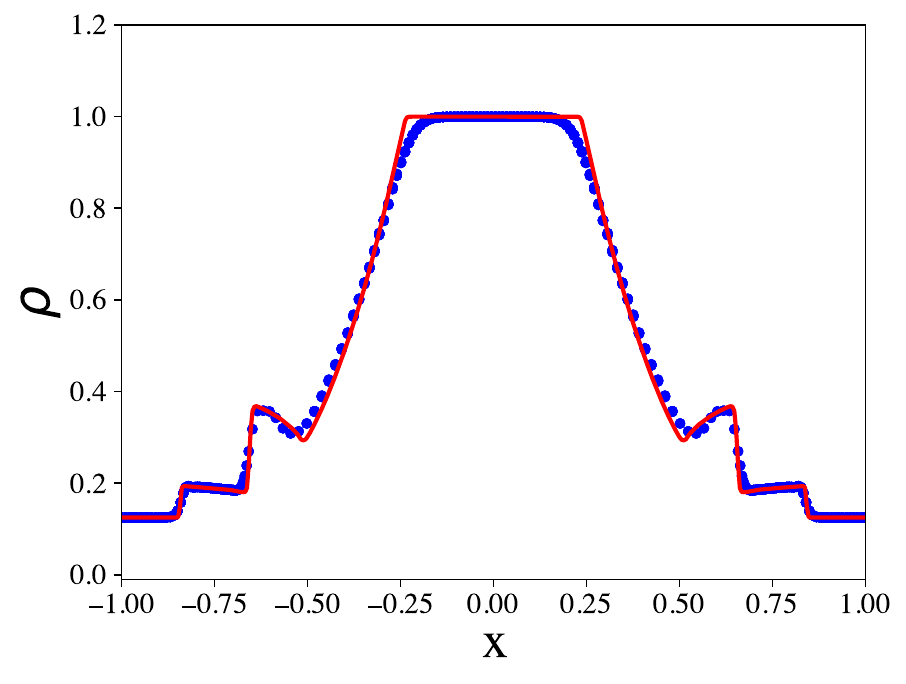} 
   \includegraphics[width=0.333\textwidth]{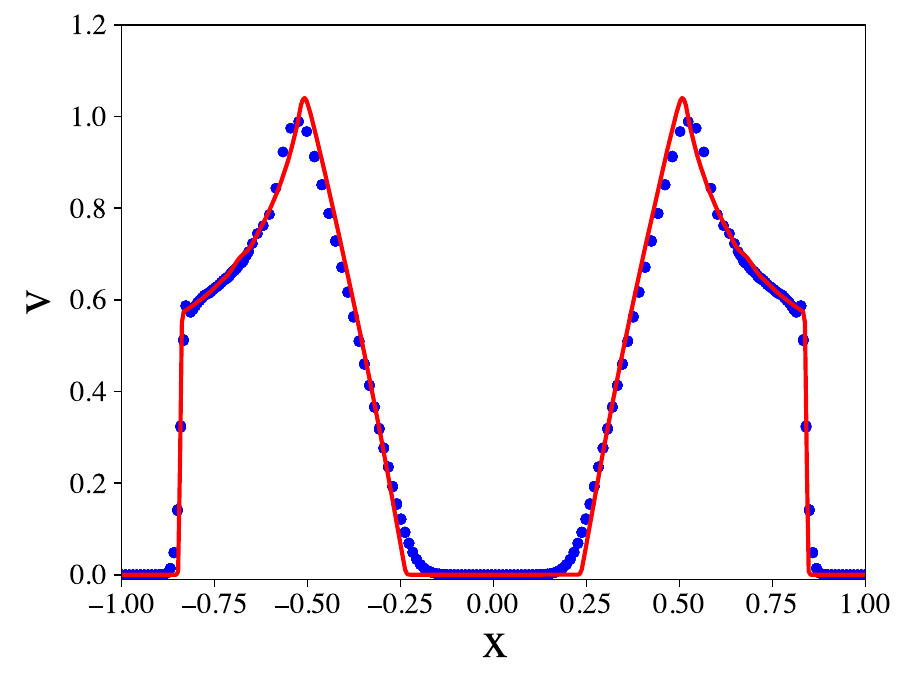} 
   \includegraphics[width=0.333\textwidth]{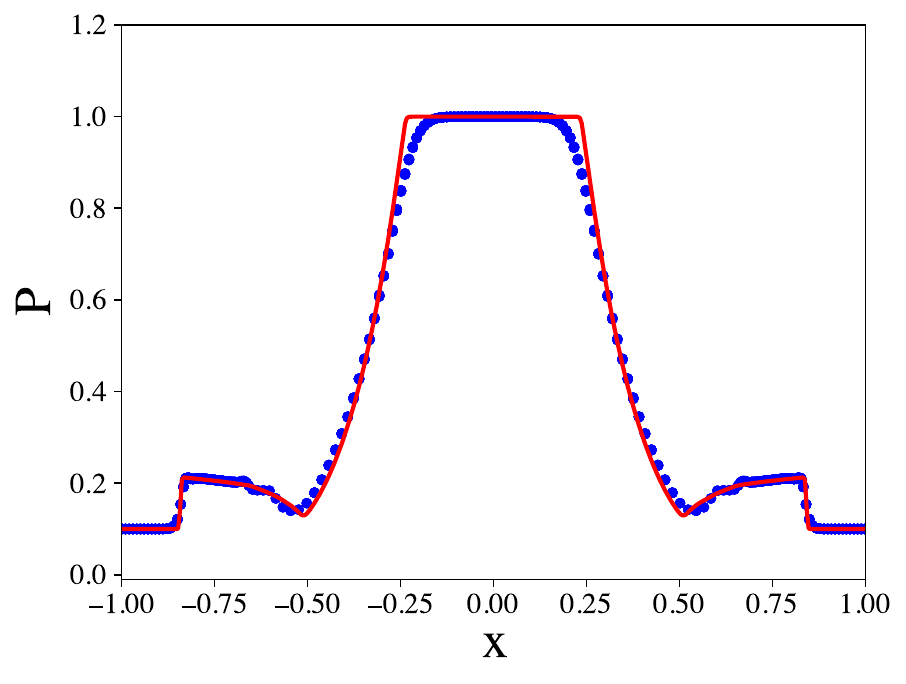} 
   }
    \caption{Spherical blast wave problem 1: density, velocity and pressure with a strip around the x axis. 
    For this simulation, $200^3$ particles and the vanAlbada limiter were used, and the reference solution (red line) was obtained by the 
    Eulerian-weighted average flux method ($400^3$ grid cells, Toro 1999).}
   \label{fig:Riemann1_Toro_comp}
\end{figure*}
\subsection{Spherical blast wave 1}
\label{subsec:Riemann1}
We start with a three-dimensional shock-tube-type problem.
We choose the same parameters as \cite{toro09}  (apart from a shift of the origin): 
the computational domain is $[-1,1]^3$ and the initial conditions are chosen as:
\be
(\rho,\vec{v},P)=
 \left\{
\begin{array}{l}
(1.000,0,0,0,1.0) \quad {\rm for \; \; r < 0.5}\\
(0.125,0,0,0,0.1) \quad {\rm else.}
\end{array}
\right.
\ee
The solution exhibits a spherical shock wave, a spherical contact surface traveling in the same direction, and a spherical rarefaction wave traveling toward the origin. As initial conditions, we simply placed the $200^3$ particles on a cubic lattice within $[-1,1]^3$, together with the surrounding "frozen" particles as the boundary condition. \\
We show in Fig.~\ref{fig:Riemann1_colour} the values of density, velocity and pressure for the 
\texttt{vanAlbada limiter}. Despite the initial setup on a cubic lattice, the results are practically perfectly spherically symmetric.
The results for the different limiters are merely identical in this case, only on very close inspection, one finds a reminiscence of the grid structure in the case of the \texttt{vanLeerMC limiter}.
In Fig.~\ref{fig:Riemann1_Toro_comp} we show our particle results in a strip around the $x-$axis ($|y| < 0.018, |z| < 0.018$) compared with a $400^3$ grid cell calculation with the Eulerian weighted average flux method \cite{toro09}. Overall, we find excellent agreement. This figure can be compared to Fig. ~13 of \cite{rosswog20a}, where the same test was performed at the same resolution but with the \Ma code using matrix inversion gradients and an artificial viscosity prescription that also uses slope-limited reconstruction. For more details on the latter method, see the section "Matrix Inversion Method I" in \cite{rosswog20a}. Although both approaches show excellent agreement with the reference solution, the RPK approach of this paper yields a sharper resolved
contact discontinuity.
\begin{figure*}
    \hspace*{-0.7cm}\includegraphics[width=1.05\textwidth]{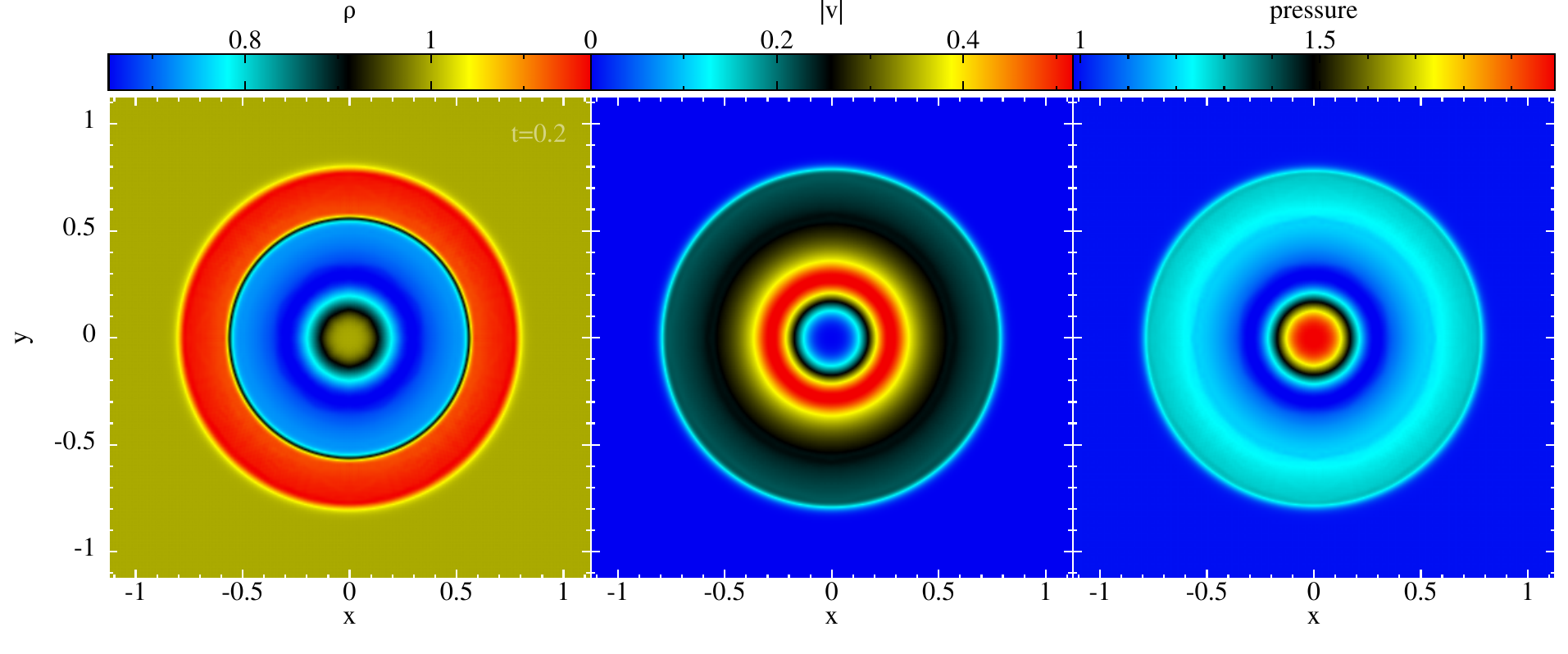} 
    \caption{Spherical blast wave problem 2: density, velocity and pressure, vanAlbada limiter.} 
   \label{fig:Riemann2_colour}
\end{figure*}

\subsection{Spherical blast wave 2}
\label{subsec:Riemann2}
In a second spherical blast wave problem \cite{toro09} we start from  
\be
(\rho,\vec{v},P)=
 \left\{
\begin{array}{l}
(1.0,0,0,0,2.0) \quad {\rm for \; \; r< 0.5}\\
(1.0,0,0,0,1.0) \quad {\rm else}
\end{array}
\right.
\ee
and place again $200^3$ particles in the same straightforward way as
in the first blast wave problem.
\\
\begin{figure*}
   \centering
   \centerline{
   \includegraphics[width=0.333\textwidth]{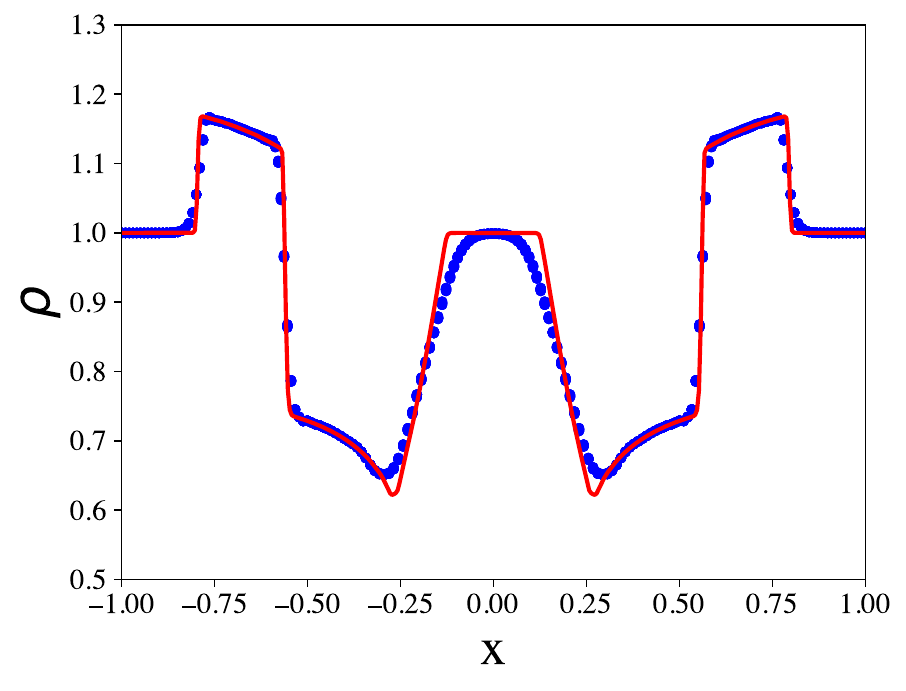} 
   \includegraphics[width=0.333\textwidth]{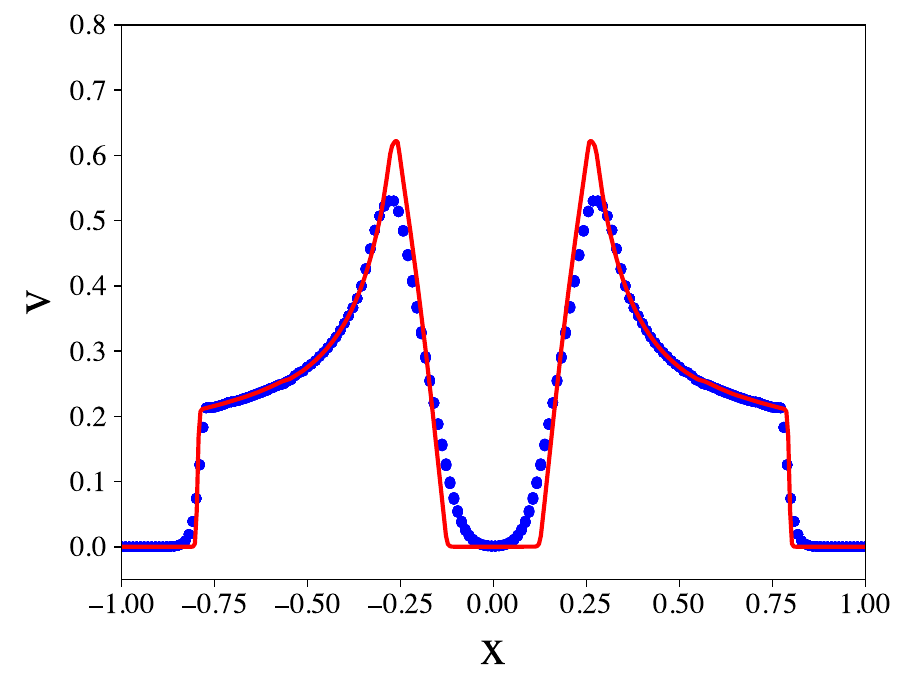} 
   \includegraphics[width=0.333\textwidth]{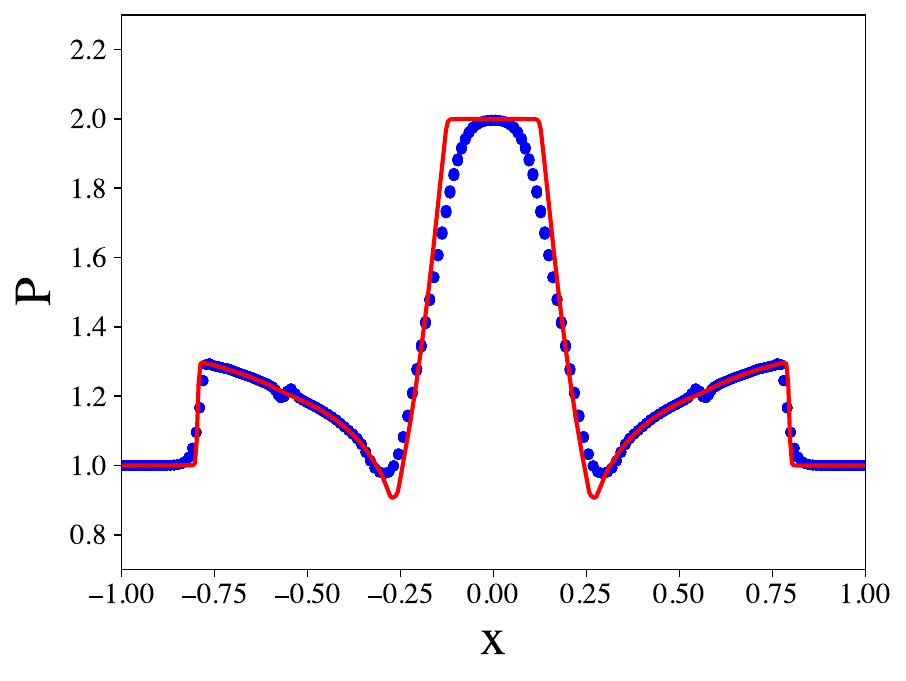} 
   }
    \caption{Spherical blast wave problem 2: density, velocity and pressure with a strip around the x axis. 
    For this simulation, $200^3$ particles and the vanAlbada limiter were used; the reference solution (red line) was obtained by the 
    Eulerian-weighted average flux method ($400^3$ grid cells, Toro 1999).}
   \label{fig:Riemann2_Toro_comp}
\end{figure*}
We show the numerical solution of this test in Fig.~\ref{fig:Riemann2_colour}. Again, deviations from sphericity are minute, and the agreement with the reference solution (Eulerian weighted average flux method with $400^3$ grid cells \cite{toro09}), see Fig.~\ref{fig:Riemann2_Toro_comp}, is excellent. In Fig.~\ref{fig:Riemann2_comparison}, we zoom in on the density distribution to illustrate the effect of the different slope limiters.  Overall, the agreement is very good, but as expected, the \texttt{minmod limiter} is the most diffusive one. The \texttt{vanLeerMC limiter} captures best the edges of the rarefaction wave, but at the price of a small overshoot at the shock front.
The results can again be compared to the  "Matrix Inversion method I" (see Figs.~14 and 15 in \cite{rosswog20a}). Overall, the results are very similar, but the RPK-hydrodynamics does not show a density overshoot in the central region as in the older result.
\begin{figure}
   \centering
   \centerline{
   \includegraphics[width=0.5\textwidth]{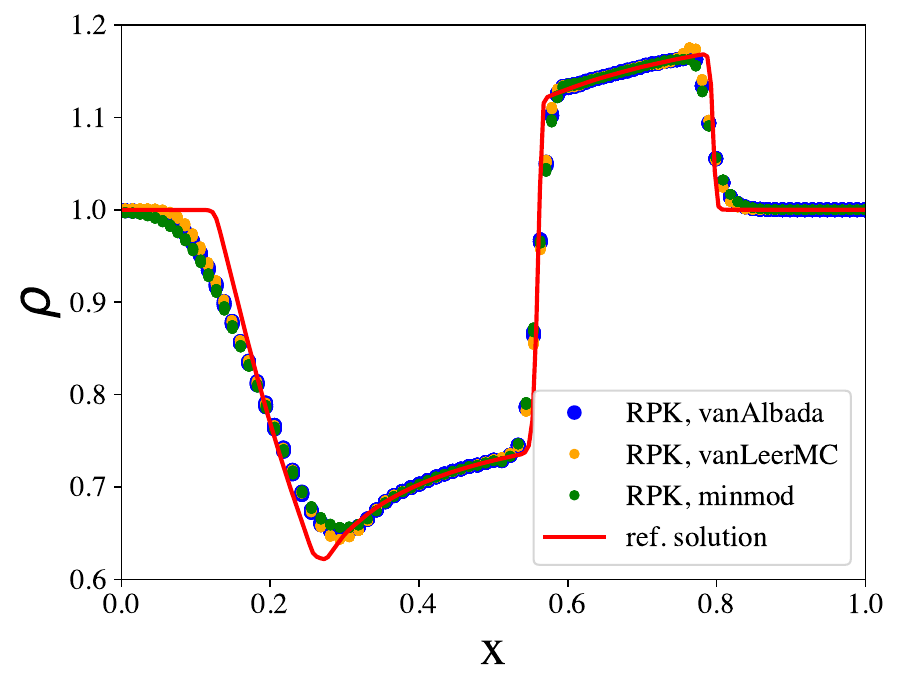} }
    \caption{Spherical blast wave problem 2 (detail): shown is the comparison of the densities for different slope limiters. }
   \label{fig:Riemann2_comparison}
\end{figure}

\subsection{Sedov blast}
\label{subsec:Sedov}
The Sedov-Taylor explosion test, a strong initial point-like explosion expanding into a low-density environment, has an analytic self-similarity 
solution \cite{sedov59,taylor50}.  For an explosion energy $E$ and a density of the ambient medium $\rho$, the blast wave propagates after 
a time $t$ to the radius $r(t)= \beta (E t^2/\rho)^{1/5}$, where $\beta$ depends on the adiabatic exponent of the gas ($\approx 1.15$ in 3D 
for the $\Gamma=5/3$ we use).  In the strong explosion limit, the density jumps in the shock front by a factor of 
$\rho_2/\rho_1= (\Gamma + 1)/(\Gamma-1)= 4$, where the numerical value refers to our chosen value of $\Gamma$. Behind the shock, the 
density drops rapidly and finally vanishes at the center of the explosion.\\
To set up the test numerically, we distribute $256^3$ SPH particles according to a {\em Centroidal Voronoi Tessellation} (CVT) 
 [-0.5,0.5]$\times$[-0.5,0.5]$\times$[-0.5,0.5]. Although this produces already reasonably good initial conditions, they can be further 
 improved by additional sweeps according to our "Artificial Pressure Method" (APM) \cite{rosswog20a}. The main idea of the APM is to 
 push each particle to a position where it minimizes its density error. Practically, this is achieved via the following steps: i) start from a 
 trial particle distribution, ii) measure the density at every particle  via Eq.~(\ref{eq:dens_sum}), iii)  calculate the error in the density 
 compared to a desired density profile, iv) from this error construct an artificial pressure and v) use this artificial pressure in an equation 
 very similar to the hydrodynamic momentum equation to push the particles in a direction where their error decreases. After many such 
 iteration steps, the particles end up in locations where they optimally approximate the desired density profile. 
 Here, we need a uniform density distribution, but the method also works very well for complicated density profiles; see, for example, 
 Fig. 3 in \cite{rosswog20a}. Even if the differences in the particle distributions are hard to see by eye, they still improve the outcome 
 of the Sedov test. We therefore use 500 of such APM sweeps here.
Once the particle distribution is settled, we assign masses so that their density is $\rho=1$. 
This is done in an iterative way where we first assign a guess value
for the masses and then measure the resulting density via Eq.~(\ref{eq:dens_sum})
and subsequently correct the particle masses. The iteration is stopped once
the density agrees everywhere to better than 0.5\% with the desired value.
The energy $E= 1$ is spread across
a very small initial radius $R$, and is distributed entirely as internal energy; the specific internal energy $u$ of the particles outside of $R$ is entirely negligible
($10^{-10}$ of the central $u$). For the initial radius $R$ we choose twice
the interaction radius of the innermost SPH particle. 
Boundaries play no role in this test as long as the blast does not interact with them.
We therefore place "frozen" particles around the computational volume as boundary particles.\\
\begin{figure*}
   \centering
   \centerline{
   \includegraphics[width=\textwidth]{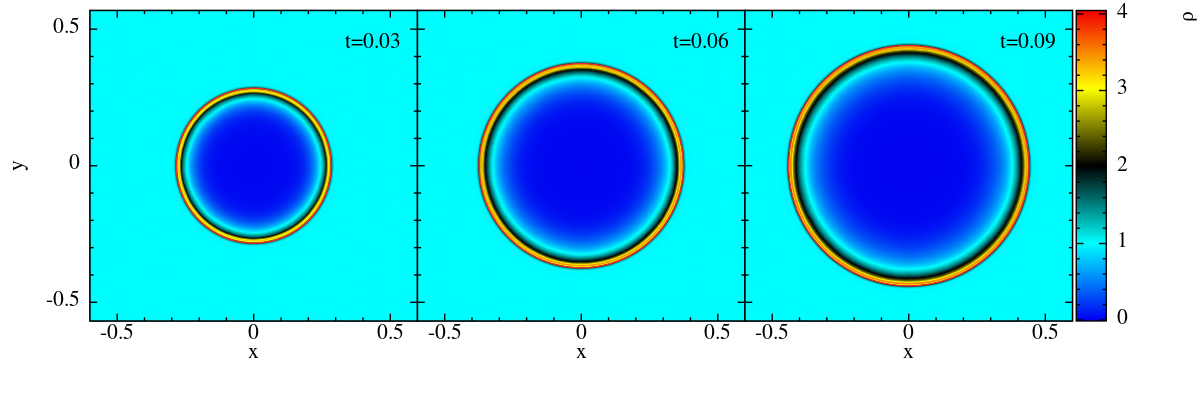} }
    \caption{Density evolution ($xy$-pane) in a Sedov blast wave test with the \texttt{vanAlbada slope limiter}. }
   \label{fig:Sedov_evolution}
\end{figure*}
In Fig.~\ref{fig:Sedov_evolution}, we show the density evolution (cut through $xy$-pane) of the case with the \texttt{vanAlbada limiter}. 
No deviation from spherical symmetry is visible, and the numerical solution for the shock  agrees very well with the exact solution 
(the hard-to-see black line that separates the background density (cyan) from the shocked matter). The density as a function of radius 
is shown in Fig.~\ref{fig:Sedov_comparison}, where each panel shows the result for one limiter. As in the tests before, and some of the
later tests, the \texttt{vanAlbada limiter} performs best. Although here \texttt{minmod} performs similarly, the \texttt{vanLeerMC limiter} does
not seem to provide enough dissipation and leads to a noisier post-shock region and an overshoot at the shock front. Again, these results 
can be compared against the \Ma results and -taken at face value- show slightly better agreement with the exact solution. However, this 
may be, at least in part, because in \cite{rosswog20a} we used a Wendland kernel with 300 neighbors for the \Ma result, while here we 
used only 220 neighbor particles.
\begin{figure*}
   \centering
   \centerline{
   \includegraphics[width=12cm]{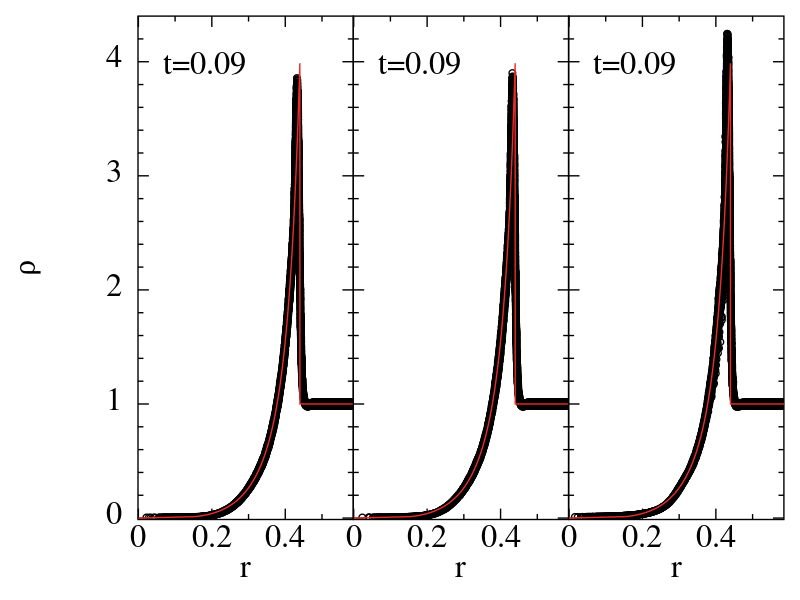} }
    \caption{Impact of the slope limiter on a Sedov blast wave: the result for \texttt{minmod} is shown in the left panel, for \texttt{vanAlbada} in the middle
    and \texttt{vanLeerMC} in the right panel.}
   \label{fig:Sedov_comparison}
\end{figure*}
\subsection{Banded Kelvin-Helmholtz instability}
\label{subsec:KeHe}
Kelvin-Helmholtz (KH) shear instabilities occur in a broad range of environments, e.g. in terrestrial clouds, in mixing processes in novae \cite{casanova11}, 
the amplification of magnetic fields in neutron star mergers \cite{price06,giacomazzo14,kiuchi15} or planetary atmospheres
\cite{johnson14}, to name just a few examples. Traditional versions of SPH have been shown to struggle with weakly triggered KH instabilities 
\cite{agertz07,mcnally12}, although many recent studies with more sophisticated numerical methods yielded very good results \cite{frontiere17,rosswog20a,sandnes24}.
We focus here on a test setup in which traditional SPH has been shown to fail, even at a rather high resolution  in 2D, see \cite{mcnally12}. 
We follow the latter paper (similar setups were used in \cite{frontiere17} and \cite{sandnes24}), but we use the full 3D code and set up the "2D" test as a thin 3D slice with 
$N \times N \times 20$ particles (referred to as $"N^2"$), For simplicity, the particles are initially placed on a cubic lattice. 
Periodic boundary conditions are obtained by placing appropriate particle copies outside of the "core" volume. 
The test is initialized as:
\be
\rho(y)=
 \left\{
\begin{array}{l}
\rho_1 - \rho_m e^{(y - 0.25)/\Delta} \quad {\rm for \; \; 0.00 \le y < 0.25}\\
\rho_2 + \rho_m e^{(0.25 - y)/\Delta} \quad {\rm for \; \; 0.25 \le y < 0.50}\\
\rho_2 + \rho_m e^{(y - 0.75)/\Delta} \quad {\rm for \; \; 0.50 \le y < 0.75}\\
\rho_1 - \rho_m e^{(0.75 - y)/\Delta} \quad {\rm for \; \; 0.75 \le y < 1.00}\\
\end{array}
\right.
\ee
where $\rho_1= 1$, $\rho_2= 2$, $\rho_m= (\rho_1 - \rho_2)/2$ and $\Delta= 0.025$.
The velocity is set up as
\be
v_x(y)=
 \left\{
\begin{array}{l}
v_1 - v_m e^{(y - 0.25)/\Delta} \quad {\rm for \; \; 0.00 \le y < 0.25}\\
v_2 + v_m e^{(0.25 - y)/\Delta} \quad {\rm for \; \; 0.25 \le y < 0.50}\\
v_2 + v_m e^{(y - 0.75)/\Delta} \quad {\rm for \; \; 0.50 \le y < 0.75}\\
v_1 - v_m e^{(0.75 - y)/\Delta} \quad {\rm for \; \; 0.75 \le y < 1.00}\\
\end{array}
\right.
\ee
with $v_1$= 0.5, $v_2= -0.5$, $v_m= (v_1-v_2)/2$ and a small velocity perturbation in $y$-direction is introduced as $v_y=  0.01 \sin(2\pi x/\lambda)$ with the perturbation wave length $\lambda= 0.5$. In the linear regime, a Kelvin-Helmholtz instability grows in the incompressible limit on a characteristic time scale of 
\be
\tau_{\rm KH}= \frac{(\rho_1+\rho_2) \lambda}{\sqrt{\rho_1 \rho_2} |v_1-v_2|},
\ee
with $\tau_{\rm KH}\approx 1.06$ for the chosen parameters.
For our tests, we chose again a polytropic equation of state with
exponent $\Gamma=5/3$. \\
\begin{figure*}
   \hspace*{-0.2cm}\includegraphics[width=1.05\textwidth]{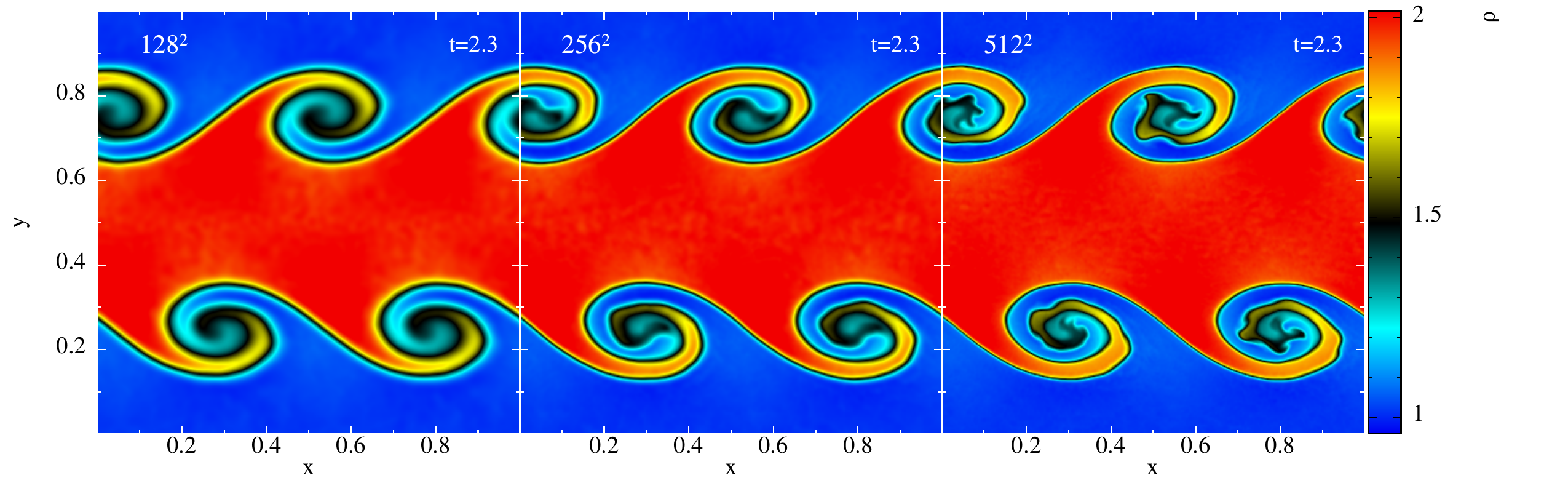} 
    \caption{Kelvin-Helmholtz test with  \texttt{vanAlbada limiter} for different resolutions at t=2.3 ($\approx 2.2 \tau_{\rm KH}$). }
   \label{fig:KeHe_vA_resolution}
\end{figure*}
\begin{figure*}
   \centerline{
   \includegraphics[width=1.05\textwidth]{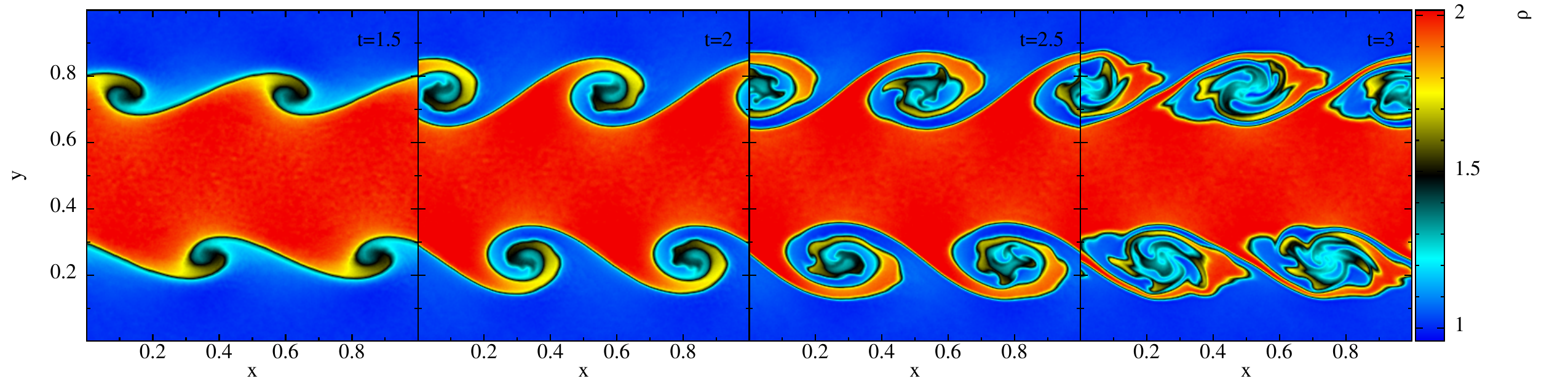} }
    \caption{Kelvin-Helmholtz test ($512^2$) with  vanAlbada limiter. }
   \label{fig:KeHe512_vA}
\end{figure*}
\begin{figure*}
   \includegraphics[width=1.05\textwidth]{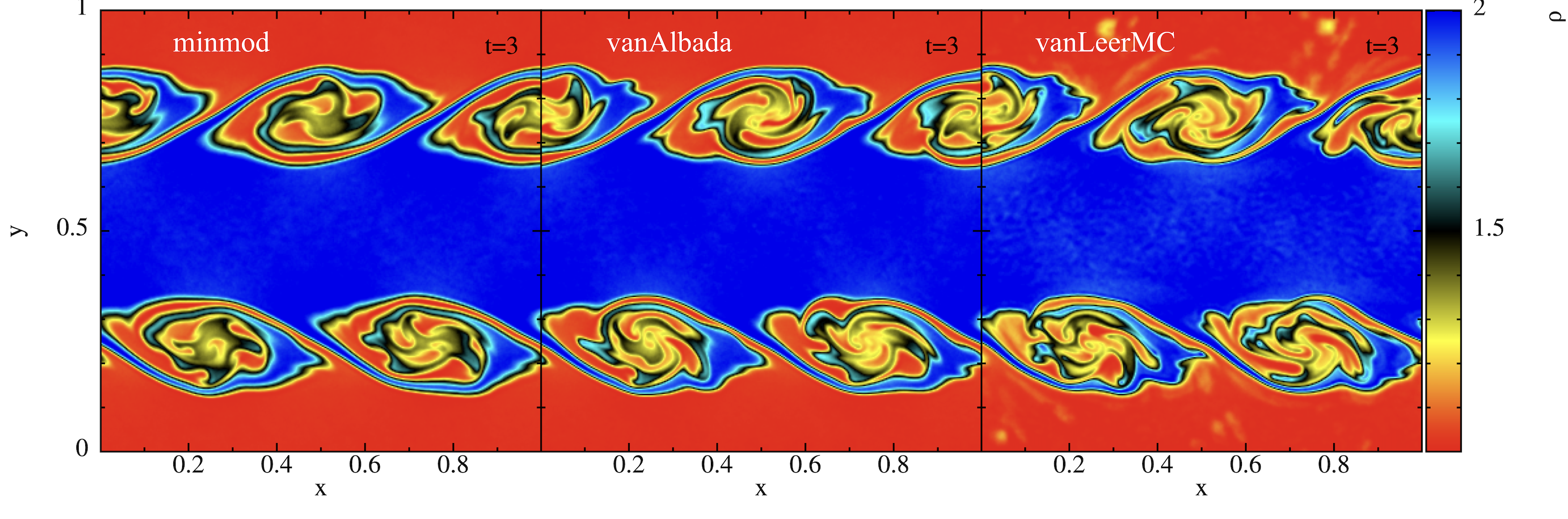}
    \caption{Kelvin-Helmholtz test ($512^2$) at t= 3 with  different slope limiters. }
   \label{fig:KeHe512_comparison}
\end{figure*}
\begin{figure*}
   \centering
   \centerline{
   \includegraphics[width=0.5\textwidth]{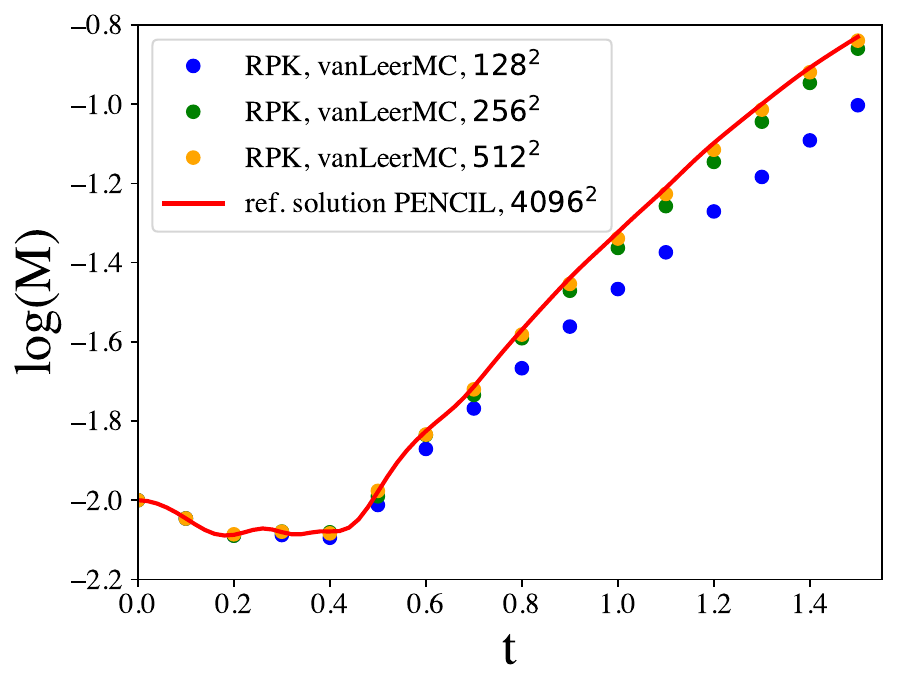}
   \includegraphics[width=0.5\textwidth]{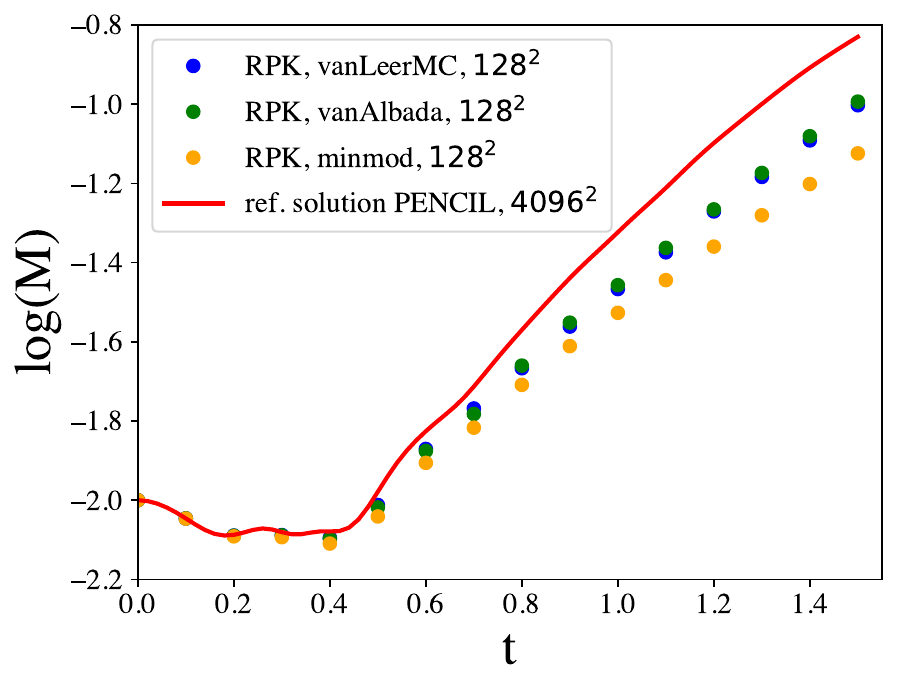}
   }
    \caption{Growth of the Kelvin-Helmholtz instability as a function of resolution (\texttt{vanLeerMC limiter}, left)
    and for a fixed, low resolution ($128^2$) for different slope limiters.}
   \label{fig:KH_growth}
 \end{figure*}
\begin{figure*}
   \centering
   \centerline{
   \includegraphics[width=0.5\textwidth]{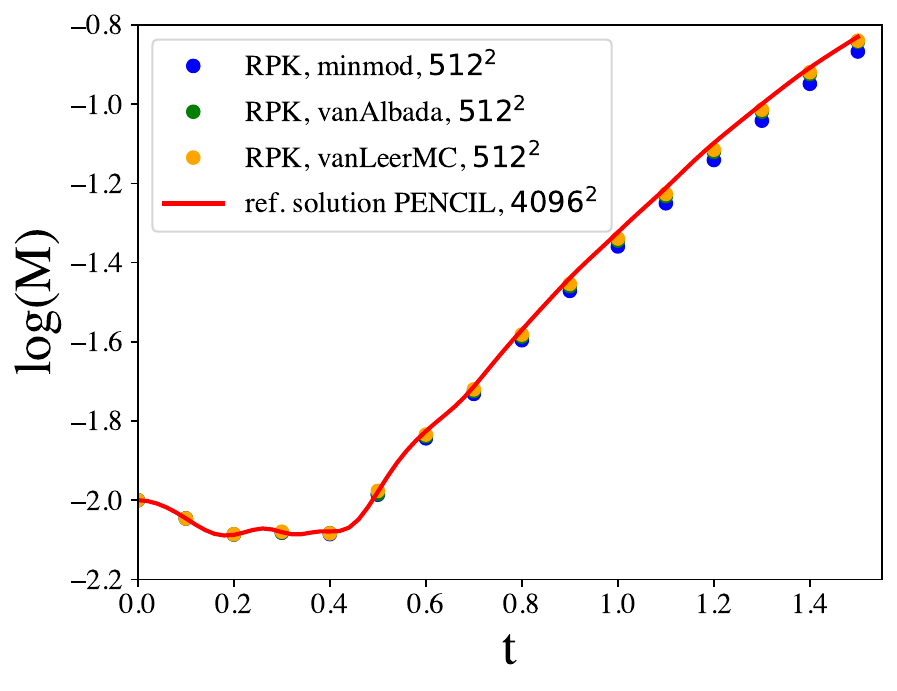}
   \includegraphics[width=0.5\textwidth]{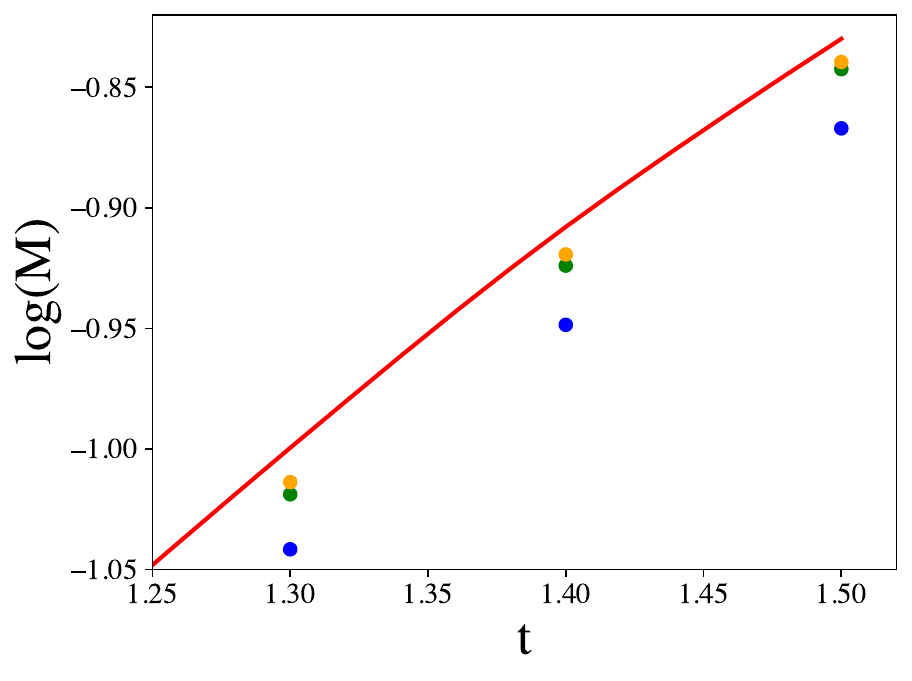}
   }
    \caption{The growth of the Kelvin-Helmholtz instability for different
      slope limiters (at $512^2$) is shown in the left panel, a zoom-in is
      shown on the right.}
   \label{fig:KH_growth_v2}
 \end{figure*}
We show in Fig.~\ref{fig:KeHe_vA_resolution} the density at $t= 2.3$ ($\approx 2.2 \; \tau_{\rm KH}$) for different resolutions. 
Even at the lowest resolution of the $128^2$ particles, we see healthy growth that is already very similar to that of the better-resolved cases.
The evolution of the $512^2$ case is shown in Fig.~\ref{fig:KeHe512_vA}.
In Fig.~\ref{fig:KeHe512_comparison}, we show the density at t=3 for the different limiters.
As expected, the \texttt{minmod limiter} is the most dissipative one, and the \texttt{vanLeerMC limiter} is once again the least dissipative.
Similarly to the Sedov test case, however, one may wonder whether the \texttt{vanLeerMC limiter} allows for enough dissipation, since 
the results seem somewhat noisy.\\ 
For comparison, traditional SPH implementations struggle with this only weakly triggered instability ($v_y= 0.01$), see, for example, Fig. 9 
of \cite{mcnally12}, where, even at a resolution of 512$^2$ particles, the instability either hardly grows (for the cubic spline kernel, label "Ne512") 
or much too slowly (for the quintic spline kernel, label "No512"). In Fig.~\ref{fig:KH_growth}, left panel, we show the mode growth (calculated 
exactly as in \cite{mcnally12}) for three different resolutions, all using the \texttt{vanLeerMC limiter}. Their growth rates are compared with a 
high-resolution reference solution ($4096^2$ cells) obtained by the PENCIL code \cite{brandenburg02}. 
Even our low-resolution case with $128^2$ particles is reasonably
close to the reference solution. To quantify the agreement with the reference solution, we calculate the quantity
  \be
  \mathcal{D}\equiv \frac{1}{N}\sqrt{\sum_p^N (M_p^{\rm ref} - M_p)^2}
  \ee
where $M$ is the mode growth exactly calculated as in \cite{mcnally12} for all our data points $p$ and the superscript ``ref'' denotes the corresponding value of the reference solution. For the simulations shown
in Fig.~\ref{fig:KH_growth} $\mathcal{D}$ is $2.84 \times 10^{-2}$ for
$128^2$, $7.05 \times 10^{-3}$ for $256^2$ and $2.65 \times 10^{-3}$ for
$512^2$. In the right panel of Fig.~\ref{fig:KH_growth}, we show the impact of
the slope limiter in the example of the lowest-resolution runs. 
The $\mathcal{D}$-values for the right panel (all $128^2$) are
$4.37 \times 10^{-2}$ for \texttt{minmod}, $2.70 \times 10^{-2}$ for
\texttt{vanAlbada} and $2.84 \times 10^{-2}$ for
\texttt{vanLeerMC}.
  In Fig.~\ref{fig:KH_growth_v2} we show the growth rates for the
  highest resolution ($512^2$) again for the different limiters (right
  panel is a zoom-in of the left one). The corresponding
  $\mathcal{D}$-values are $6.92 \times 10^{-3}$ for \texttt{minmod}, $3.61 \times 10^{-3}$ for
\texttt{vanAlbada} and $2.65 \times 10^{-3}$ for
\texttt{vanLeerMC}. So the most diffusive \texttt{minmod limiter} grows slowest, while the
  growth rates of \texttt{vanAlbada} and \texttt{vanLeerMC} are very
  similar (\texttt{vanAlbada} having a small advantage at $128^2$
  while \texttt{vanLeerMC} performs slightly better at $512^2$) .
\begin{figure*}
   \centering
   \centerline{
   \includegraphics[width=\textwidth]{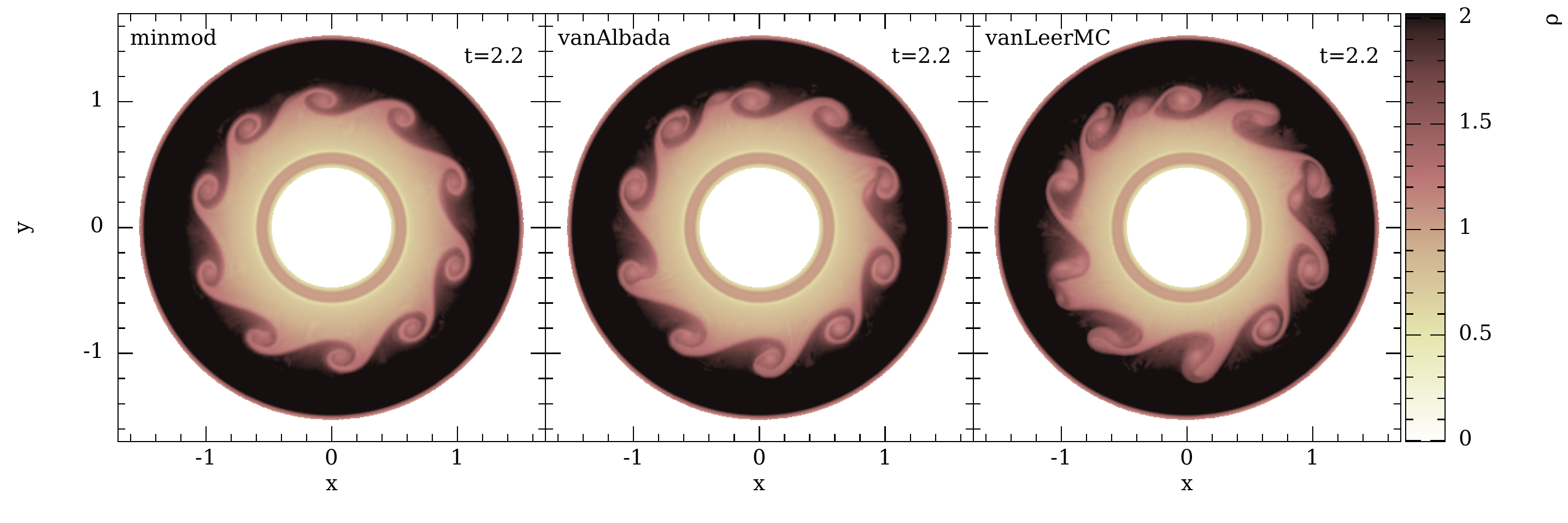}}
    \caption{Density in cylindrical Kelvin-Helmholtz test for the different slope limiters.}
   \label{fig:cylKeHe}
\end{figure*}

\subsection{Cylindrical Kelvin-Helmholtz instability}
\label{subsec:cylKeHe}
As a variant of the Kelvin-Helmholtz instability, we also perform a test in cylindrical symmetry, similar to \cite{duffell16}. To this end, we place particles from a radius
of $R_{\rm BD1}= 0.5$ to $R_{\rm BD2}= 1.5$. For the region with $r < 1$, we use
\be
\rho=1 \quad \omega=2 \quad P= 4 + 2 r^2,
\ee and we also use
\be
\rho=2 \quad \omega=1 \quad P= 5 +  r^2,
\ee
 where $\omega$ is the angular frequency. In addition, we impose a small radial velocity perturbation on the interface given by
\be
v_{\rm r }= v_0 \cos(10 \; \varphi) \;  e^{-\frac{(r-1)^2}{2 \sigma_0^2}},
\ee
where $\varphi$ is the azimuthal angle, $v_0= 0.02$ and $\sigma_0= 0.1$.\\
In our initial setup, we placed 1.95 million particles in a uniform, close-packed lattice so that the particles
have the above properties. Again, we use the full 3D code to simulate a slice with 20 layers of particles in the $z$-direction, we do not allow for motion in the $z$-direction, and we enforce the
above velocities within $R_{\rm BD1}$ and $R_{\rm BD1} + 0.05$ and $R_{\rm BD2} - 0.05$ and $R_{\rm BD2}$ as boundary conditions.\\
We show in Fig.~\ref{fig:cylKeHe} the density at $t=2.2$, again for our three limiters. There are small artefacts at the inner and outer boundaries because of our simple treatment of the
boundary conditions, but in all cases, we see healthy growing Kelvin-Helmholtz instabilities with nicely winding billows. For the most diffusive limiter (\texttt{minmod}), we see 
essentially only the triggered modes growing. In contrast, in the other two cases, "parasitic" modes have also developed, which have been triggered by the granularity of our closely-packed particles shearing against each other.

\subsection{Rayleigh-Taylor instability}
\label{subsec:RaTa}
\begin{figure*}
   \centering
   \centerline{
   \includegraphics[width=\textwidth]{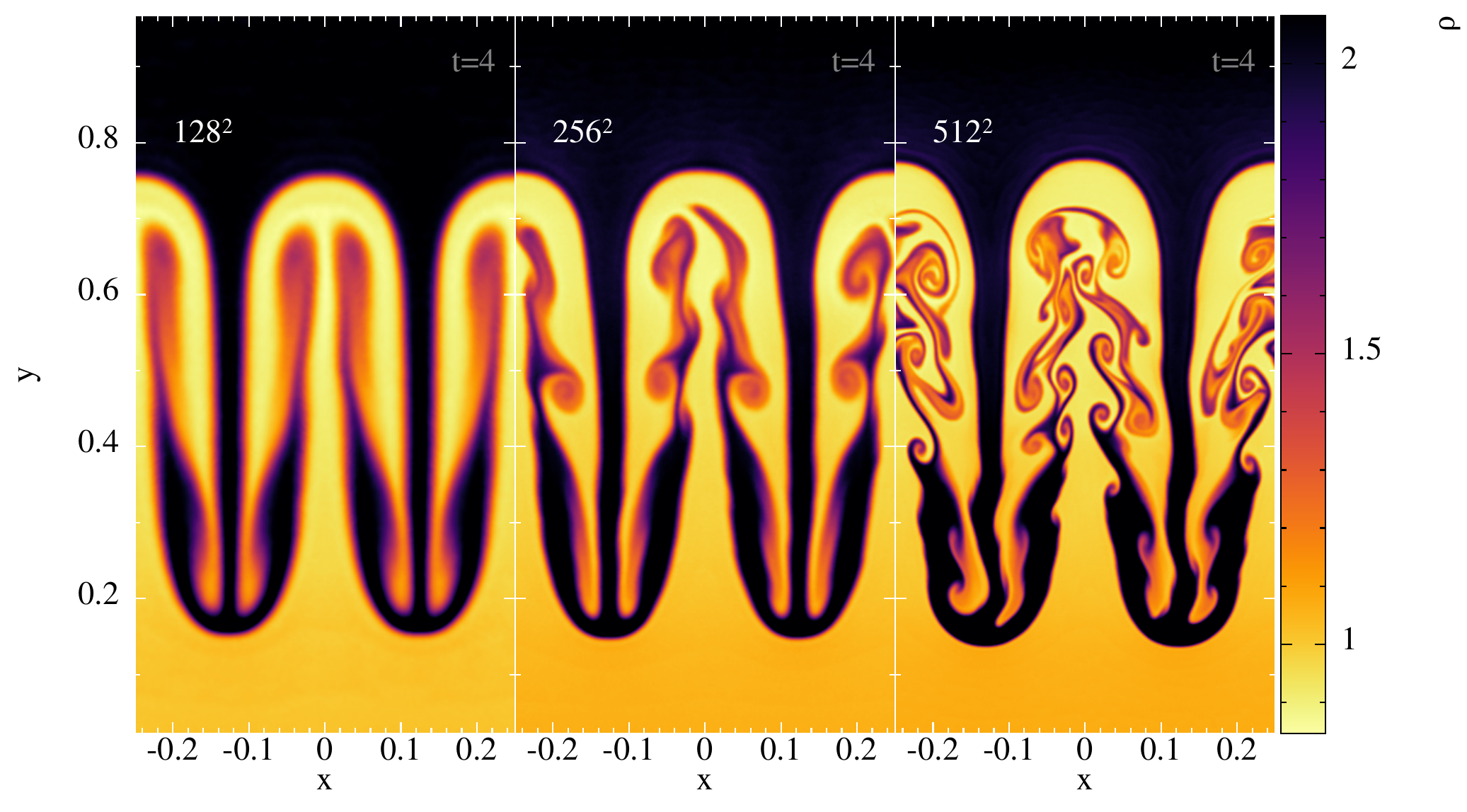}}
    \caption{Rayleigh-Taylor instability test (at t=4) with \texttt{vanAlbada limiter} at different resolution : $128^2$ (left), $256^2$ (middle) and $512^2$ (right).}
   \label{fig:RaTa_res}
\end{figure*}
%
\begin{figure*}
   \centering
   \centerline{
   \includegraphics[width=1.1\textwidth]{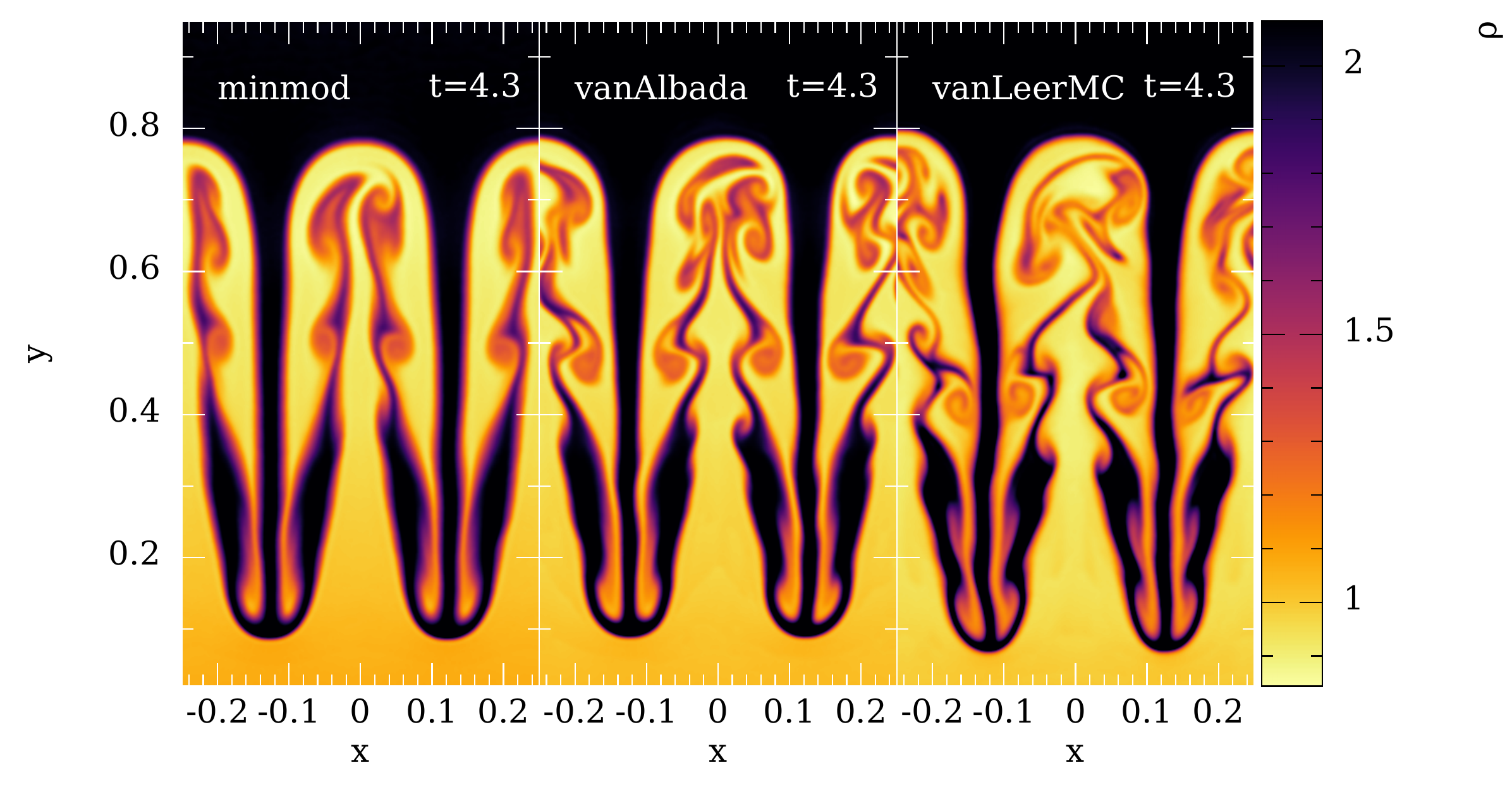}}
    \caption{Rayleigh-Taylor instability test (at t= 4.3) with $256^2$ particles and different slope limiters.}
   \label{fig:RaTa_limiter}
\end{figure*}
The Rayleigh-Taylor instability is a standard probe of the subsonic growth of a small perturbation.
In its simplest form, a density layer $\rho_t$ rests on top of a layer with density $\rho_b < \rho_t$ in a constant acceleration field, e.g. due to gravity. 
While the denser fluid sinks down, it develops a characteristic "mushroom-like" pattern.  Simulations with traditional SPH implementations have 
shown only retarded growth or even complete suppression of instability \cite{abel11,saitoh13}.\\
We again adopt a quasi-2D setup and use the full 3D code for the evolution.
We place the particles on a cubic lattice in the $xy$-domain $[-0.25,0.25] \times [0,1]$ and use 20 layers of particles in the $z$-direction, and also place 20 layers of particles as boundaries
around this core region. The properties of particles below $y=0$ are "frozen" at the values of the initial conditions, for particles with $y_a > 1$, we multiply a damping factor
\be
\xi= \exp\left\{-\left(\frac{y_a - 1}{0.05}\right)^2\right\}
\ee 
to the temporal derivatives so that any evolution in this upper region is strongly suppressed. We apply
periodic boundaries in $x-$direction at $x= \pm 0.25$. Similar to \cite{frontiere17} we use 
$\rho_t=2$, $\rho_b=1$, a constant acceleration $\vec{g}= -0.5 \hat{e}_y$ and
\be
\rho(y)= \rho_b + \frac{\rho_t-\rho_b}{1+\exp[-(y-y_t)/\Delta]}
\ee
with transition width $\Delta=0.025$ and transition coordinate $y_t=0.5$. We apply a small velocity perturbation to the interface
\be
v_y(x,y)= \delta v_{y,0} [1 + \cos(8\pi x)][1 + \cos(5\pi(y-y_t))]
\ee
for $y$ in $[0.3,0.7]$ with an initial amplitude $\delta v_{y,0}=0.025$, and we use a polytropic
equation of state with exponent $\Gamma=1.4$. The equilibrium pressure profile
is given by
\be
P (y) = P_0 -  g \rho(y) [y - y_t]
\ee
with $P_0= \rho_t/\Gamma$, so that the sound speed is near unity in the transition region.
\\
We show in Fig.~\ref{fig:RaTa_res} the results for different resolutions at $t=4$ for the \texttt{vanAlbada limiter}. At all resolutions, the instability grows and reaches similar $y$ values, but, of course, with finer detailed structures for higher resolution.
In Fig.~\ref{fig:RaTa_limiter}, we show the medium resolution case for the different limiters, again confirming the hierarchy of dissipation levels with \texttt{minmod} being the most and \texttt{vanLeerMC} being the least dissipative limiter.
\begin{figure*}
   \centering
   \includegraphics[width=\textwidth]{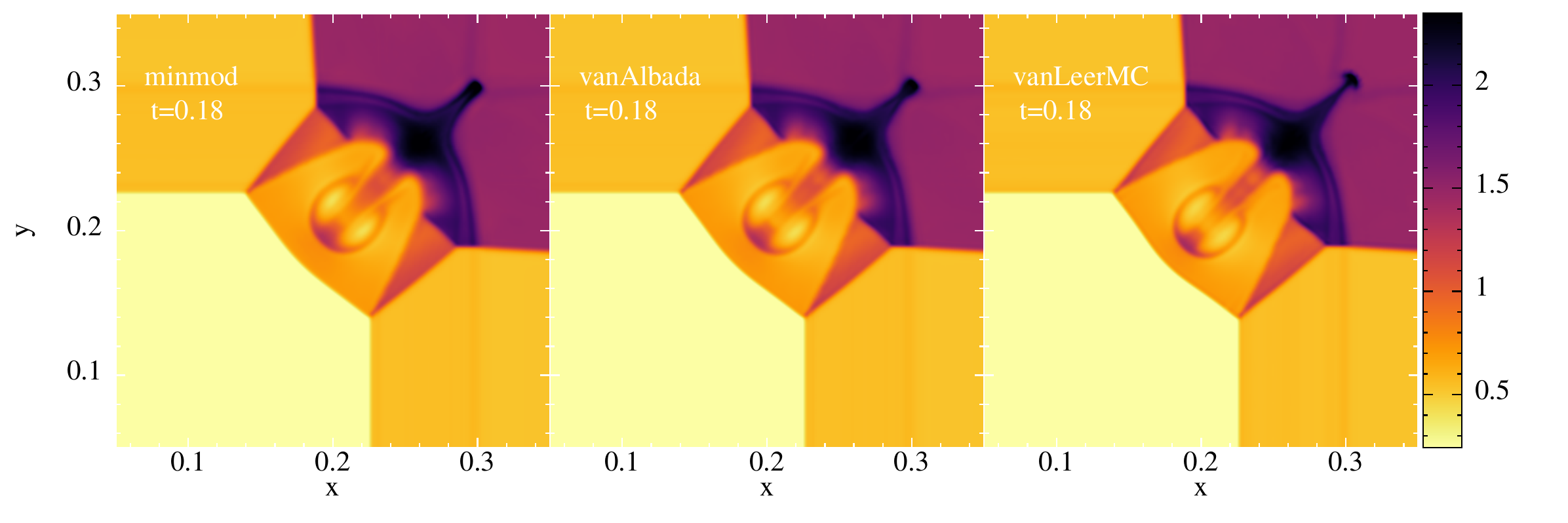}
   \includegraphics[width=\textwidth]{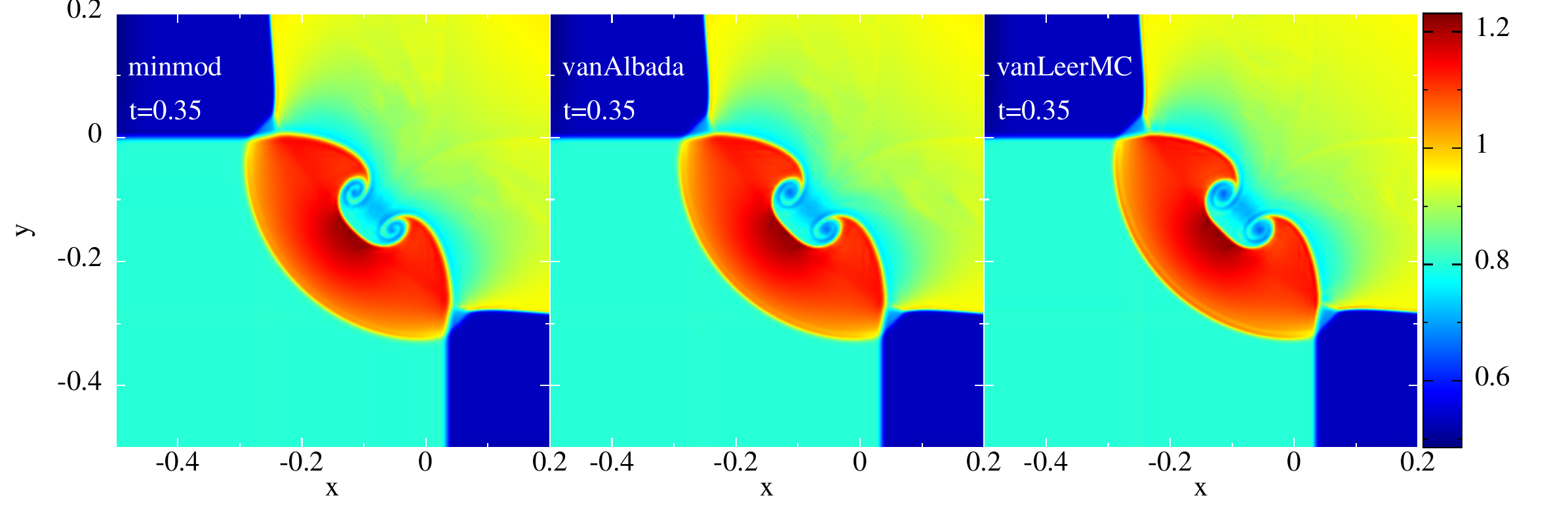}
   \includegraphics[width=\textwidth]{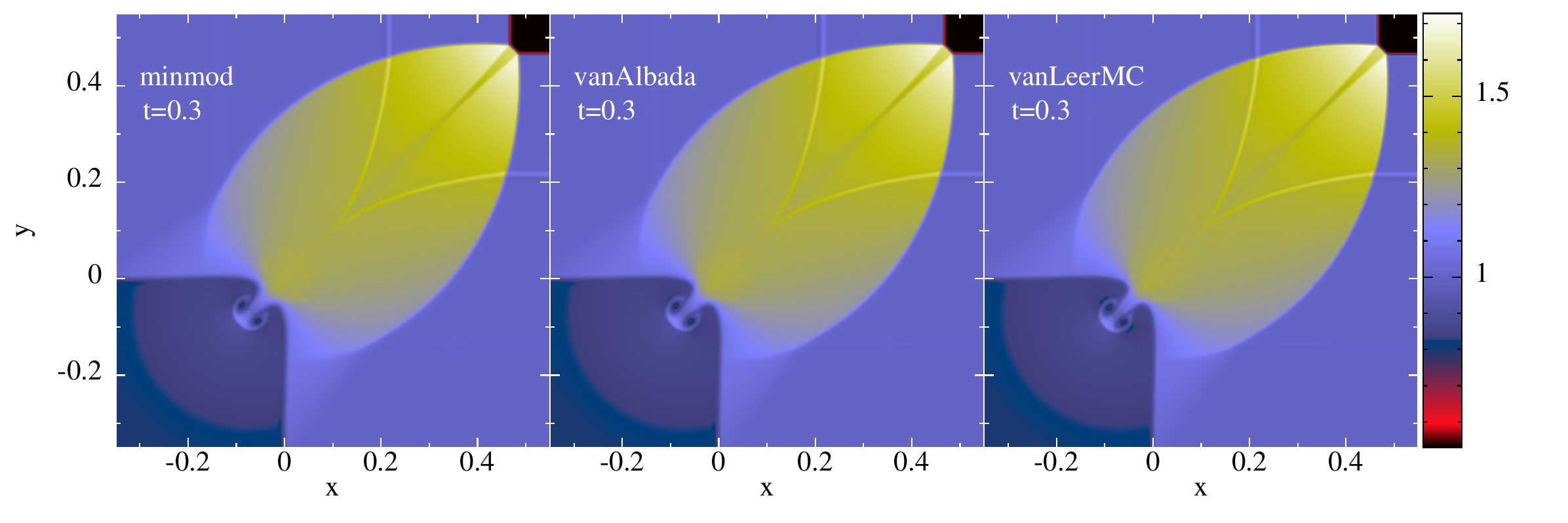}
    \caption{Schulz-Rinne test 1 (first row), test 2 (second row), and test 3 (third row), color-coded is the density. The first column
    each time refers to \texttt{minmod}, the second to \texttt{vanAlbada} and the third one to \texttt{vanLeerMC}.}
   \label{fig:SchuRi_all}
\end{figure*}

\subsection{Complex shocks with vorticity creation}
\begin{table}
\caption{Initial data for the Schulz-Rinne-type 2D Riemann problems with vorticity creation.}
\begin{tabular}{ l c | c | c | c | c | c | c |}
\hline
 & & SR1; contact point: $(0.3,0.3)$ & &\\
  \hline	
  \hline		
  variable & NW & NE & SW & SE \\ \hline
  $\rho$ &  0.5323 &  1.5000  & 0.1380 &  0.5323 \\
  $v_x$ &  1.2060 &  0.0000   & 1.2060 & 0.0000  \\
  $v_y$ &   0.0000 & 0.0000   & 1.2060 & 1.2060 \\
  $P$    &   0.3000 & 1.5000   & 0.0290 & 0.3000    \\
  \hline  
   & & SR2; contact point: $(0.0,0.0)$ & &\\
  \hline	
  \hline		
  variable & NW & NE & SW & SE \\ \hline
  $\rho$ &  0.5313 &  1.0000 & 0.8000  &  0.5313 \\
  $v_x$ &   0.8276 &  0.1000 & 0.1000 & 0.1000  \\
  $v_y$ &   0.0000 & 0.0000  & 0.0000  &  0.7276\\
  $P$    &   0.4000  & 1.0000  & 0.4000 & 0.4000    \\
  \hline  
   & & SR3; contact point: $(0.0,0.0)$ & &\\
  \hline	
  variable & NW & NE & SW & SE \\ \hline
  $\rho$ &  1.0000 &  0.5313 & 0.8000  &  1.000 \\
  $v_x$ &   0.7276 &  0.0000 & 0.0000 & 0.0000  \\
  $v_y$ &   0.0000 & 0.0000  & 0.0000  &  0.7262\\
  $P$    &   1.0000  & 0.4000  & 1.0000 & 1.0000    \\
  \hline  
\end{tabular}
\label{tab:SchulzRinne}
\end{table}
Schulz-Rinne \cite{schulzrinne93a} suggested a set of very challenging 2D benchmark tests.
The tests are constructed so that four constant states meet at one corner, and the initial values are chosen so that an elementary wave, either a shock, a rarefaction, or a contact discontinuity, appears at each interface.  
During subsequent evolution, complex wave patterns
emerge for which exact solutions are not known. These tests are considered very challenging benchmarks for multidimensional hydrodynamic codes \cite{schulzrinne93a,lax98,kurganov02,liska03}. Such tests have rarely been shown for SPH. We are only aware of the work by \cite{puri14}, who show results for one such shock test
in a study of Godunov SPH with approximate Riemann solvers and as benchmarks for the \Ma code \cite{rosswog20a}.\\
Here, we investigate three such configurations. 
We simulate, as before, a 3D slice thick enough so that the midplane is unaffected by edge effects (we use 20 particle layers in $z$-direction).  
We place particles on a cubic lattice so that 300 x 300 particles are within  $[x_c-0.3,x_c+0.3] \times [y_c-0.3,y_c+0.3]$, 
where $(x_c,y_c)$ is the contact point of the quadrants, and we use
a polytropic exponent $\Gamma=1.4$ in all tests. We refer to these Schulz-Rinne type problems as SR1-SR3 and give their initial parameters for each quadrant in Tab. \ref{tab:SchulzRinne}. 
These test problems correspond to configurations 3, 11, and 12 in the labeling convention of \cite{kurganov02}. We smooth the initial conditions through $\langle A \rangle_a= \sum_b \left[(m_b/\rho_b) A_b W_{ab}\right]/\left[\sum_b (m_b/\rho_b)  W_{ab}\right]$ so that they have a smoothness that is consistent with the evolution code.\\
For all three tests, we find very good agreement with the results published in the literature \cite{schulzrinne93a,lax98,kurganov02,liska03,rosswog20a}. Comparing the performance of the different limiters for test SR1 (first row in Fig.~\ref{fig:SchuRi_all}), we see only minor differences. All cases nicely produce the shock surfaces and all show the "mushroom-like" structure near (0.2,0.2) with only small differences. The structure around (0.3,0.3), however, does show some differences: it looks "mushroom-like" for the least dissipative case (\texttt{vanLeerMC}), but not
very much so for the more dissipative limiters. Also, test SR2 agrees overall well with the results in the literature, but here
the "mushroom-like" structures are different: the curls show more windings for lower dissipation, that is, the smallest number
for \texttt{minmod} and the highest for \texttt{vanLeerMC}. The SR3 test also agrees well with the results in the literature, here
 the results for the different limiters are virtually identical.

\section{Conclusion}
\label{sec:summary}
Here, we have presented and explored a new formulation of particle hydrodynamics. We started from a common SPH formulation and enhanced it with Roe's approximate Riemann solver to solve Riemann problems between interacting particle pairs.
More specifically, we replaced the mean values of the pressures and (projected) velocities with the contact discontinuity values
$P^\ast$ and $v^\ast$. We identified the terms in the Roe solver that are responsible for the dissipation. In these terms, we replaced the particle properties with the values found by slope-limited reconstruction from both sides to the interparticle midpoint.
We use kernels that are designed to enforce consistency relations and which, by construction, recover constant and linear functions to machine precision, called "reproducing kernels". We have carefully scrutinized our corresponding implementation based on a "glass-like" particle distribution, and we compare the function and gradient reproduction to the standard SPH approximation. We find that the errors in this test
are at least nine orders of magnitude lower for the RPK than for the standard SPH approach. These reproducing kernels are used both for the hydrodynamic gradients and in the reconstruction process.\\
We have put our new formulation to the test in many challenging benchmarks, ranging from shocks over instabilities to
vorticity-creating shock tests designed by Schulz-Rinne. We find very good results in all tests and, where comparable, we find slightly better but overall very similar results to the \Ma code.  We have also explored three different slope limiters in the reconstruction: \texttt{minmod, vanAlbada}, and \texttt{vanLeerMC}. Although \texttt{minmod} is an overall robust choice, it leads to more dissipation than appears necessary. The \texttt{vanLeerMC limiter}, in contrast, produces the least dissipation of the three limiters, but maybe not enough in the Sedov blast wave test, where it leads to substantially more post-shock noise than the other limiters, and an overshoot of the density. Of the three explored limiters, our clear favorite is the \texttt{vanAlbada} limiter, which performs well in all tests, but it may be worth exploring further limiters in the future.
Although the results presented are very encouraging, more work needs to be done. This may include further studies of robustness and accuracy, a careful comparison with modern SPH formulations,  and the inclusion of additional physics. However, these explorations are left for future work.

%

\section*{Acknowledgements}
The simulations for this paper have been performed on the facilities of
 North-German Supercomputing Alliance (HLRN), and at the SUNRISE 
 HPC facility supported by the Technical Division at the Department of 
 Physics, Stockholm University, and on the HUMMEL2 cluster funded 
 by the Deutsche Forschungsgemeinschaft  (498394658). Special thanks 
 go to Mikica Kocic (SU),  Thomas Orgis and Hinnerk St\"uben (both UHH) for their 
 excellent support.
 Some  plots have been produced with the software \texttt{splash}
 \cite{price07d}.

 \section*{Funding statement}
 The author has been supported by the Swedish Research Council (VR) under 
grant number 2020-05044, by the research environment grant
``Gravitational Radiation and Electromagnetic Astrophysical
Transients'' (GREAT) funded by the Swedish Research Council (VR) 
under Dnr 2016-06012, by the Knut and Alice Wallenberg Foundation
under grant Dnr. KAW 2019.0112,  by the Deutsche 
Forschungsgemeinschaft (DFG, German Research Foundation) under 
Germany's Excellence Strategy - EXC 2121 ``Quantum Universe'' - 390833306 
and by the European Research Council (ERC) Advanced 
Grant INSPIRATION under the European Union's Horizon 2020 research 
and innovation programme (Grant agreement No. 101053985).

\section*{Conflicts of interest}
The author has no conflicts of interest for the present study.

\section*{Ethics Approval}
Not applicable.

\section*{Data availability}
The data underlying this article will be shared on reasonable request to the corresponding author.

\nocite{*}
\bibliographystyle{unsrt}
\bibliography{SC_RPK}

\begin{thebibliography}{10}

\bibitem{lucy77}
L.B. Lucy.
\newblock A numerical approach to the testing of the fission hypothesis.
\newblock {\em The Astronomical Journal}, 82:1013, 1977.

\bibitem{gingold77}
R.~A. {Gingold} and J.~J. {Monaghan}.
\newblock {Smoothed particle hydrodynamics: theory and application to
  non-spherical stars.}
\newblock {\em Monthly Notices of the Royal Astronomical Society},
  181:375--389, November 1977.

\bibitem{xu21}
X.~{Xu}, Y.-L. {Jiang}, and P.~{Yu}.
\newblock {SPH simulations of 3D dam-break flow against various forms of the
  obstacle}.
\newblock {\em Ocean Engineering}, 229:108978, 2021.

\bibitem{rahimi22}
M.~N. {Rahimi} and G.~{Moutsanidis}.
\newblock {A smoothed particle hydrodynamics approach for phase field modeling
  of brittle fracture}.
\newblock {\em Computer Methods in Applied Mechanics and Engineering},
  398:115191, August 2022.

\bibitem{monaghan12a}
J.~J. {Monaghan}.
\newblock {Smoothed Particle Hydrodynamics and Its Diverse Applications}.
\newblock {\em Annual Review of Fluid Mechanics}, 44:323--346, January 2012.

\bibitem{rosswog22a}
Stephan {Rosswog}.
\newblock {Modelling astrophysical fluids with particles}.
\newblock In Dmitry {Bisikalo}, Dmitri {Wiebe}, and Christian {Boily}, editors,
  {\em The Predictive Power of Computational Astrophysics as a Discover Tool},
  volume 362 of {\em IAU Symposium}, pages 382--397, January 2023.

\bibitem{monaghan05}
J.~J. {Monaghan}.
\newblock {Smoothed particle hydrodynamics}.
\newblock {\em Reports on Progress in Physics}, 68:1703--1759, August 2005.

\bibitem{rosswog09b}
S.~Rosswog.
\newblock Astrophysical smooth particle hydrodynamics.
\newblock {\em New Astronomy Reviews}, 53:78--104, 2009.

\bibitem{springel10a}
V.~{Springel}.
\newblock {Smoothed Particle Hydrodynamics in Astrophysics}.
\newblock {\em ARAA}, 48:391--430, September 2010.

\bibitem{price12a}
D.~J. {Price}.
\newblock {Smoothed particle hydrodynamics and magnetohydrodynamics}.
\newblock {\em Journal of Computational Physics}, 231:759--794, February 2012.

\bibitem{rosswog15c}
S.~{Rosswog}.
\newblock {SPH Methods in the Modelling of Compact Objects}.
\newblock {\em Living Reviews of Computational Astrophysics (2015)}, 1, October
  2015.

\bibitem{rosswog21a}
S.~{Rosswog} and P.~{Diener}.
\newblock {SPHINCS\_BSSN: a general relativistic smooth particle hydrodynamics
  code for dynamical spacetimes}.
\newblock {\em Classical and Quantum Gravity}, 38(11):115002, June 2021.

\bibitem{rosswog23a}
Stephan {Rosswog}, Francesco {Torsello}, and Peter {Diener}.
\newblock {The Lagrangian Numerical Relativity code SPHINCS\_BSSN\_v1.0}.
\newblock {\em Front. Appl. Math. Stat.}, 9, October 2023.

\bibitem{bambi25}
S.~{Rosswog} and P.~{Diener}.
\newblock Sphincs\_bssn: Numerical relativity with particles.
\newblock In Cosimo Bambi, Yosuke Mizuno, Swarnim Shashank, and Feng Yuan,
  editors, {\em New Frontiers in GRMHD Simulations}. Springer, Singapore, 2025.

\bibitem{morris97}
J.P. Morris and J.J. Monaghan.
\newblock A switch to reduce sph viscosity.
\newblock {\em J. Comp. Phys.}, 136:41, 1997.

\bibitem{rosswog00}
S.~{Rosswog}, M.~B. {Davies}, F.-K. {Thielemann}, and T.~{Piran}.
\newblock {Merging neutron stars: asymmetric systems}.
\newblock {\em A\&A}, 360:171--184, August 2000.

\bibitem{cullen10}
L.~{Cullen} and W.~{Dehnen}.
\newblock {Inviscid smoothed particle hydrodynamics}.
\newblock {\em MNRAS}, 408:669--683, October 2010.

\bibitem{rosswog15b}
S.~{Rosswog}.
\newblock {Boosting the accuracy of SPH techniques: Newtonian and
  special-relativistic tests}.
\newblock {\em MNRAS}, 448:3628--3664, March 2015.

\bibitem{rosswog20b}
S.~{Rosswog}.
\newblock {A Simple, Entropy-based Dissipation Trigger for SPH}.
\newblock {\em ApJ}, 898(1):60, July 2020.

\bibitem{frontiere17}
N.~{Frontiere}, C.~D. {Raskin}, and J.~M. {Owen}.
\newblock {CRKSPH - A Conservative Reproducing Kernel Smoothed Particle
  Hydrodynamics Scheme}.
\newblock {\em Journal of Computational Physics}, 332:160--209, March 2017.

\bibitem{rosswog20a}
S.~{Rosswog}.
\newblock {The Lagrangian hydrodynamics code MAGMA2}.
\newblock {\em MNRAS}, 498(3):4230--4255, aug 2020.

\bibitem{sandnes24}
Thomas~D. {Sandnes}, Vincent~R. {Eke}, Jacob~A. {Kegerreis}, Richard~J.
  {Massey}, Sergio {Ruiz-Bonilla}, Matthieu {Schaller}, and Luis F.~A.
  {Teodoro}.
\newblock {REMIX SPH -- improving mixing in smoothed particle hydrodynamics
  simulations using a generalised, material-independent approach}.
\newblock {\em arXiv e-prints}, page arXiv:2407.18587, July 2024.

\bibitem{inutsuka02}
S.-I. {Inutsuka}.
\newblock {Reformulation of Smoothed Particle Hydrodynamics with Riemann
  Solver}.
\newblock {\em Journal of Computational Physics}, 179:238--267, June 2002.

\bibitem{parshikov02}
Anatoly~N. {Parshikov} and Stanislav~A. {Medin}.
\newblock {Smoothed Particle Hydrodynamics Using Interparticle Contact
  Algorithms}.
\newblock {\em Journal of Computational Physics}, 180(1):358--382, July 2002.

\bibitem{sirotkin13}
F.V. {Sirotkin} and J.K. {Yoh}.
\newblock {A Smoothed Particle Hydrodynamics method with approximate Riemann
  solvers for simulation of strong explosions}.
\newblock {\em Computers and Fluids}, 88:418--429, 2013.

\bibitem{puri14}
K.~{Puri} and P.~{Ramachandran}.
\newblock {Approximate Riemann solvers for the Godunov SPH (GSPH)}.
\newblock {\em Journal of Computational Physics}, 270:432--458, August 2014.

\bibitem{meng21}
Z.F. {Meng}, A.M. {Zhang}, P.P. {Wang}, and F.R. {Ming}.
\newblock A shock-capturing scheme with a novel limiter for compressible flows
  solved by smoothed particle hydrodynamics.
\newblock {\em Computer Methods in Applied Mechanics and Engineering},
  386:114082, 2021.

\bibitem{wendland95}
Holger Wendland.
\newblock Piecewise polynomial, positive definite and compactly supported
  radial functions of minimal degree.
\newblock {\em Advances in Computational Mathematics}, 4(1):389--396, 1995.

\bibitem{cabezon08}
R.~M. {Cabezon}, D.~{Garcia-Senz}, and A.~{Relano}.
\newblock {A one-parameter family of interpolating kernels for smoothed
  particle hydrodynamics studies}.
\newblock {\em Journal of Computational Physics}, 227:8523--8540, October 2008.

\bibitem{dehnen12}
W.~{Dehnen} and H.~{Aly}.
\newblock {Improving convergence in smoothed particle hydrodynamics simulations
  without pairing instability}.
\newblock {\em MNRAS}, 425:1068--1082, September 2012.

\bibitem{garcia_senz12}
D.~{Garcia-Senz}, R.~{Cabezon}, and J.A. Escartin.
\newblock {Improving smoothed particle hydrodynamics with an integral approach
  to calculating gradients}.
\newblock {\em A \& A}, 538:A9, February 2012.

\bibitem{cabezon12a}
R.~M. {Cabezon}, D.~{Garcia-Senz}, and J.~A. {Escartin}.
\newblock {Testing the concept of integral approach to derivatives within the
  smoothed particle hydrodynamics technique in astrophysical scenarios}.
\newblock {\em A \& A}, 545:A112, September 2012.

\bibitem{antona21}
Rub{\'e}n {Antona}, Renato {Vacondio}, Diego {Avesani}, Maurizio {Righetti},
  and Massimiliano {Renzi}.
\newblock Towards a high order convergent ale\_sph scheme with efficient weno
  spatial reconstruction.
\newblock {\em Water}, 13(17):2432, 2021.

\bibitem{avesani21}
Diego {Avesani}, Michael {Dumbser}, Renato {Vacondio}, and Maurizio {Righetti}.
\newblock {An alternative SPH formulation: ADER-WENO-SPH}.
\newblock {\em Computer Methods in Applied Mechanics and Engineering},
  382:113871, August 2021.

\bibitem{vergnaud23}
A.~{Vergnaud}, G.~{Oger}, and D.~{Le Touz{\'e}}.
\newblock {Investigations on a high order SPH scheme using WENO
  reconstruction}.
\newblock {\em Journal of Computational Physics}, 477:111889, March 2023.

\bibitem{antona24}
Rub{\'e}n {Antona}, Renato {Vacondio}, Diego {Avesani}, Maurizio {Righetti},
  and Massimiliano {Renzi}.
\newblock {A WENO SPH scheme with improved transport velocity and consistent
  divergence operator}.
\newblock {\em Computational Particle Mechanics}, 11(3):1221--1240, June 2024.

\bibitem{benmoussa99}
B.~{Ben Moussa}, N.~{Lanson}, and J.P. {Vila}.
\newblock Convergence of meshless methods for conservation laws: Applications
  to euler equations.
\newblock {\em International Series of Numerical Mathematics}, 29, 1999.

\bibitem{vila99}
J.P. {Vila}.
\newblock On particle weighted methods and smooth particle hydrodynamics.
\newblock {\em Mathematical Models and Methods in Applied Science},
  02:161--209, 1999.

\bibitem{hietel00}
Dietmar Hietel, Konrad Steiner, and Jens Struckmeier.
\newblock A finite-volume particle method for compressible flows.
\newblock {\em Mathematical Models and Methods in Applied Sciences},
  10(09):1363--1382, 2000.

\bibitem{junk03}
M.~{Junk}.
\newblock Do finite volume methods need a mesh?
\newblock {\em In: Griebel M., Schweitzer M.A. (eds) Meshfree Methods for
  Partial Differential Equations. Lecture Notes in Computational Science and
  Engineering, vol 26. Springer, Berlin, Heidelberg}, 26:223--238, 2003.

\bibitem{gaburov11}
E.~{Gaburov} and K.~{Nitadori}.
\newblock {Astrophysical weighted particle magnetohydrodynamics}.
\newblock {\em MNRAS}, 414:129--154, June 2011.

\bibitem{ramirez22}
Luis {Ramirez}, Antonio {Eiris}, Ivan {Couceiro}, Jose {Paris}, and Xesus
  {Nogueira}.
\newblock {An arbitrary Lagrangian-Eulerian SPH-MLS method for the computation
  of compressible viscous flows}.
\newblock {\em Journal of Computational Physics}, 464:111172, September 2022.

\bibitem{schulzrinne93a}
C.~W. {Schulz-Rinne}.
\newblock {Classification of the Riemann problem for two-dimensional gas
  dynamics}.
\newblock {\em SIAM Journal of Mathematical Analysis}, 24:76--88, January 1993.

\bibitem{liu03}
G.R. {Liu} and M.B. {Liu}.
\newblock {\em Smoothed Particle Hydrodynamics: A Meshfree Particle Method}.
\newblock World Scientific, 2003.

\bibitem{rosswog22b}
Stephan {Rosswog}, Peter {Diener}, and Francesco {Torsello}.
\newblock {Thinking Outside the Box: Numerical Relativity with Particles}.
\newblock {\em Symmetry}, 14(6):1280, June 2022.

\bibitem{toro09}
E.~F. {Toro}.
\newblock {\em Riemann solvers and Numerical methods for fluid dynamics}.
\newblock Springer-Verlag, Berlin, 3rd ed. edition, 2009.

\bibitem{roe86}
P~L Roe.
\newblock Characteristic-based schemes for the euler equations.
\newblock {\em Annual Review of Fluid Mechanics}, 18(1):337--365, 1986.

\bibitem{vanLeer77}
Bram {van Leer}.
\newblock {Towards the Ultimate Conservative Difference Scheme. IV. A New
  Approach to Numerical Convection}.
\newblock {\em Journal of Computational Physics}, 23:276, March 1977.

\bibitem{vanAlbada82}
G.~D. {van Albada}, B.~{van Leer}, and Jr. {Roberts}, W.~W.
\newblock {A comparative study of computational methods in cosmic gas
  dynamics}.
\newblock {\em A \& A}, 108(1):76--84, April 1982.

\bibitem{liu95}
Wing~Kam {Liu}, Sukky {Jun}, and Yi~Fei {Zhang}.
\newblock {Reproducing kernel particle methods}.
\newblock {\em International Journal for Numerical Methods in Fluids},
  20(8-9):1081--1106, Apr 1995.

\bibitem{du05}
Q.~Du and D.~Wang.
\newblock The optimal centroidal voronoi tessellations and the gersho's
  conjecture in the three-dimensional space.
\newblock {\em Computers and Mathematics with Applications}, 49:1355, 2005.

\bibitem{gafton11}
E.~{Gafton} and S.~{Rosswog}.
\newblock {A fast recursive coordinate bisection tree for neighbour search and
  gravity}.
\newblock {\em MNRAS}, 418:770--781, December 2011.

\bibitem{gottlieb98}
S.~{Gottlieb} and C.~W. {Shu}.
\newblock {Total variation diminishing Runge-Kutta schemes}.
\newblock {\em Mathematics of Computation}, 67:73--85, January 1998.

\bibitem{owen14}
J.~M. {Owen}.
\newblock {A compatibly differenced total energy conserving form of SPH}.
\newblock {\em Numerical Methods in Fluids}, 75:749--774, August 2014.

\bibitem{cercos_pita23}
Jose~Luis {Cercos-Pita}, Pablo~Eleazar {Merino-Alonso}, Javier
  {Calderon-Sanchez}, and Daniel {Duque}.
\newblock {The role of time integration in energy conservation in Smoothed
  Particle Hydrodynamics fluid dynamics simulations}.
\newblock {\em European Journal of Mechanics, B/Fluids}, 97:78--92, January
  2023.

\bibitem{sedov59}
L.~I. {Sedov}.
\newblock {\em Similarity and Dimensional Methods in Mechanics}.
\newblock Academic Press, New York, 1959.

\bibitem{taylor50}
G.~{Taylor}.
\newblock {The Formation of a Blast Wave by a Very Intense Explosion. I.
  Theoretical Discussion}.
\newblock {\em Proceedings of the Royal Society of London Series A},
  201:159--174, March 1950.

\bibitem{casanova11}
J.~{Casanova}, J.~{Jose}, E.~{Garcia-Berro}, S.~N. {Shore}, and A.~C. {Calder}.
\newblock {Kelvin-Helmholtz instabilities as the source of inhomogeneous mixing
  in nova explosions}.
\newblock {\em Nature}, 478:490--492, October 2011.

\bibitem{price06}
D.J. Price and S.~Rosswog.
\newblock Producing ultra-strong magnetic fields in neutron star mergers.
\newblock {\em Science}, 312:719, 2006.

\bibitem{giacomazzo14}
Bruno {Giacomazzo}, Jonathan {Zrake}, Paul~C. {Duffell}, Andrew~I. {MacFadyen},
  and Rosalba {Perna}.
\newblock {Producing Magnetar Magnetic Fields in the Merger of Binary Neutron
  Stars}.
\newblock {\em ApJ}, 809(1):39, August 2015.

\bibitem{kiuchi15}
K.~{Kiuchi}, P.~{Cerd{\'a}-Dur{\'a}n}, K.~{Kyutoku}, Y.~{Sekiguchi}, and
  M.~{Shibata}.
\newblock {Efficient magnetic-field amplification due to the Kelvin-Helmholtz
  instability in binary neutron star mergers}.
\newblock {\em Phys. Rev. D}, 92(12):124034, December 2015.

\bibitem{johnson14}
J.~R. {Johnson}, S.~{Wing}, and P.~A. {Delamere}.
\newblock {Kelvin Helmholtz Instability in Planetary Magnetospheres}.
\newblock {\em Space Science Reviews}, 184:1--31, November 2014.

\bibitem{agertz07}
O.~{Agertz}, B.~{Moore}, J.~{Stadel}, D.~{Potter}, F.~{Miniati}, J.~{Read},
  L.~{Mayer}, A.~{Gawryszczak}, A.~{Kravtsov}, {\AA}.~{Nordlund}, F.~{Pearce},
  V.~{Quilis}, D.~{Rudd}, V.~{Springel}, J.~{Stone}, E.~{Tasker},
  R.~{Teyssier}, J.~{Wadsley}, and R.~{Walder}.
\newblock {Fundamental differences between SPH and grid methods}.
\newblock {\em MNRAS}, 380:963--978, September 2007.

\bibitem{mcnally12}
C.~P. {McNally}, W.~{Lyra}, and J.-C. {Passy}.
\newblock {A Well-posed Kelvin-Helmholtz Instability Test and Comparison}.
\newblock {\em ApJS}, 201:18, August 2012.

\bibitem{brandenburg02}
A.~{Brandenburg} and W.~{Dobler}.
\newblock {Hydromagnetic turbulence in computer simulations}.
\newblock {\em Computer Physics Communications}, 147:471--475, August 2002.

\bibitem{duffell16}
Paul~C. {Duffell}.
\newblock {DISCO: A 3D Moving-mesh Magnetohydrodynamics Code Designed for the
  Study of Astrophysical Disks}.
\newblock {\em \apjs}, 226(1):2, September 2016.

\bibitem{abel11}
T.~{Abel}.
\newblock {rpSPH: a novel smoothed particle hydrodynamics algorithm}.
\newblock {\em MNRAS}, 413:271--285, May 2011.

\bibitem{saitoh13}
T.~R. {Saitoh} and J.~{Makino}.
\newblock {A Density-independent Formulation of Smoothed Particle
  Hydrodynamics}.
\newblock {\em ApJ}, 768:44, May 2013.

\bibitem{lax98}
P.~{Lax} and X.D. {Liu}.
\newblock {Solution of Two-Dimensional Riemann Problems of Gas Dynamics by
  Positive Schemes}.
\newblock {\em SIAM J. Sci. Comput}, 19:319--340, January 1998.

\bibitem{kurganov02}
A.~{Kurganov} and Eitan {Tadmor}.
\newblock {Solution of Two-Dimensional Riemann Problems for Gas Dynamics
  without Riemann Problem Solvers}.
\newblock {\em Numerical Methods for Partial Differential Equations},
  18:584--608, 2002.

\bibitem{liska03}
R.~Liska and B.~Wendroff.
\newblock Comparison of several difference schemes on 1d and 2d test problems
  for the euler equations.
\newblock {\em SIAM J. Sci. Comput.}, 25:995 -- 1017, 2003.

\bibitem{price07d}
D.~J. {Price}.
\newblock {splash: An Interactive Visualisation Tool for Smoothed Particle
  Hydrodynamics Simulations}.
\newblock {\em Publications of the Astronomical Society of Australia},
  24:159--173, October 2007.

\end{thebibliography}

\end{document}